\documentclass[12pt]{article}

\usepackage{color,amsmath,amssymb,exscale,psfrag,epsfig}
\usepackage{cite,color,url}
\usepackage[colorlinks=true
,urlcolor=blue
,anchorcolor=blue
,citecolor=blue
,filecolor=blue
,linkcolor=blue
,menucolor=blue
,linktocpage=true
,pdfproducer=medialab
]{hyperref}
\input epsf
\usepackage{soul}
\usepackage{aas_macros}

\newcommand{\m}[1]{\marginpar{{\tiny *}} }

\def\bea{\begin{eqnarray}}
\def\eea{\end{eqnarray}}

\newcommand{\lsi}{\,\raisebox{-0.13cm}{$\stackrel{\textstyle<}
{\textstyle\sim}$}\,}
\newcommand{\gsi}{\,\raisebox{-0.13cm}{$\stackrel{\textstyle> 
}
{\textstyle\sim}$}\,}

\newcommand{\rev}[1]{#1}

\catcode`\@=11
\def\lsim{\mathrel{\mathpalette\@versim<}}
\def\gsim{\mathrel{\mathpalette\@versim>}}
\def\@versim#1#2{\vcenter{\offinterlineskip
\ialign{$\m@th#1\hfil##\hfil$\crcr#2\crcr\sim\crcr } }}
\catcode`\@=12
\catcode`@=12
\begin{document}
\thispagestyle{empty}
\begin{flushright}
\end{flushright}
\vspace{0.3in}
\begin{center}
{\Large \bf {Searches for correlation between UHECR events and high-energy gamma-ray {\it Fermi}-LAT data}} \\
\vspace{0.5in}
{\bf Ezequiel \'Alvarez$^a$, Alessandro Cuoco$^b$, Nestor Mirabal$^{c,d}$, Gabrijela Zaharijas$^e$}
\vspace{0.2in} \\
{\sl $^a$ IFIBA (CONICET) Ciudad Universitaria, Pab.1, (1428) Buenos Aires, Argentina; and 
International Center for Advanced Studies (ICAS), UNSAM, Campus Miguelete, 25 de Mayo y Francia, 1650 Buenos Aires, Argentina} \\
{\sl $^b$ Institute for Theoretical Particle Physics and Cosmology (TTK), RWTH Aachen University, D-52056 Aachen,
Germany} \\
{\sl $^c$ NASA Goddard Space Flight Center, Greenbelt, MD 20771, USA} \\
{\sl $^d$ NASA Postdoctoral Program Fellow, USA} \\
{\sl $^e$ {\it 
Laboratory for Astroparticle Physics, University of Nova Gorica,
	Vipavska 13, SI-5000 Nova Gorica, Slovenia} }
\end{center}
\vspace{0.3in}

\begin{abstract}
The astrophysical sources responsible for ultra high-energy cosmic rays (UHECRs) continue to be one 
of the most intriguing mysteries in astrophysics.
We present a comprehensive search for correlations between 
high-energy ($\gsi 1$ GeV) gamma-ray events from the {\it Fermi} Large Area Telescope (LAT) 
and UHECRs ($\gsi 60$ EeV) detected by the Telescope
Array and the Pierre Auger Observatory. 
We perform two separate searches. 
First, we conduct a standard cross-correlation analysis
between the arrival directions of 148 UHECRs and 360 gamma-ray sources 
in the Second Catalog of Hard Fermi-LAT sources (2FHL). 
Second, we search for a possible
correlation between UHECR directions and unresolved {\it Fermi}-LAT
  gamma-ray emission.
For the latter, we use three different methods: a stacking technique with both a model-dependent and model-independent background estimate,
and a cross-correlation function analysis. {\rev We also test for
  statistically significant excesses in gamma rays from signal regions centered
on  Cen A and the Telescope Array hotspot.} 
No significant correlation is found in any of the analyses performed, except 
a weak ($\lsi 2\sigma$) hint of signal with the correlation function method on scales $\sim 1^\circ$.
Upper limits on the flux of possible  power-law gamma-ray  sources
of UHECRs are derived. 

\end{abstract}

\vspace*{10mm}
\noindent {\footnotesize E-mail:
{\tt \href{mailto:sequi@df.uba.ar}{sequi@df.uba.ar},
\href{mailto:cuoco@physik.rwth.aachen.de}{cuoco@physik.rwth.aachen.de},
\href{mailto:nestor.r.mirabalbarrios@nasa.gov}{nestor.r.mirabalbarrios@nasa.gov},\\
\href{mailto:gabrijela.zaharijas@ung.si}{gabrijela.zaharijas@ung.si}}}


\section{Introduction}

The astrophysical sources responsible for the acceleration of Ultra High Energy Cosmic Rays (UHECRs) remain a mystery despite many decades of research \cite{Linsley:1963km}. A number of viable theoretical candidates able to reach energies above $10^{19}$ eV exist, 
including active galactic nuclei (AGN) \cite{Hillas:1985is} and gamma-ray bursts (GRBs) \cite{Waxman:1995vg}. Unfortunately, the trajectory of charged UHECRs can be deflected by intervening magnetic fields. As a result, tracing the origin of a particular UHECR back to its original source in the sky is rather not trivial. Cross correlation between UHECRs and catalogues of specific classes of objects has been the typical route to search for clues. This has been done for the Supergalactic plane \cite{Stanev:1995my}, IRAS galaxies \cite{Cuoco:2005yd,Waxman:1996hp},  AGN \cite{Abraham:2007si},  gamma-ray sources \cite{2010MNRAS.405L..99M}  and  IceCube high-energy neutrinos  \cite{Moharana:2015nxa,Aartsen:2015dml}. Searches conducted so far have produced no significant correlation.  

Here we look for a correlation between UHECR  and high-energy ($\gsi 1$ GeV) 
gamma-ray events collected by the Large Area Telescope (LAT) on board the {\it Fermi} space observatory \cite{Atwood:2009ez}.
We use UHECR data from the Pierre Auger Observatory (PAO),
located in Argentina, and the Telescope Array (TA) located in Utah,
thus providing nearly all-sky coverage.
Most of the proposed energetic UHECR sources are expected 
to be gamma-ray emitters, making such search  well motivated. 
Furthermore, 
both high-energy  gamma rays and UHECRs should 
come from relatively nearby, a fact that could further enhance correlations, if present. 
In fact, UHECRs are expected to come within  $z\lsi 0.05$ for energies $\gsi 40$ EeV due
to the GZK~\cite{Greisen:1966jv,Zatsepin:1966jv} attenuation by pion production with the cosmic microwave background (CMB),
while high-energy gamma rays ($\gsi 50$ GeV) are attenuated by pair production on CMB (and its analogous infrared and radio counterparts)
which creates an effective horizon of $z\lsi 1$ .

Some authors argued that high-frequency peaked  (or extreme-frequency peaked) 
BL Lacs  gamma-ray emission might have a hadronic origin, 
producing protons of UHECR energies \cite{Tavecchio:2015cid,Boettcher:2013wxa}. 
These sources are more naturally observed above hundreds of GeVs, by ground based Atmospheric Cherenkov Telescopes (ACTs).
However, the advantage of the LAT is that it has whole sky coverage, and therefore provides an unbiased survey of these objects. In addition the sensitivity of the LAT is steadily increasing with time particularly with the newest, Pass 8 \cite{Atwood:2013rka}, event level analysis. 

We perform two different correlation studies: correlation of the UHECRs with the hard gamma-ray point sources and 
correlation with diffuse photons. 
In the first case we perform  a `standard' correlation search between UHECRs and point sources from the 2FHL catalog \cite{Ackermann:2015uya}.
The latter is based on 80 months of LAT data and the newly delivered 
Pass 8 event-level analysis \cite{Atwood:2013rka}. The increased statistics and effective area of this event class resulted in a significant increase in the sensitivity in the 50 GeV -- 2 TeV energy range, with the discovery of 360 sources. 
Searches for correlations with various source catalogues have been 
performed in the past, but here, for the first time, we will focus on the latest LAT catalog, which is specifically useful to study the high-energy range. 
For the second part of the work, we present 
a search for possible gamma-ray excesses over the isotropic diffuse gamma-ray background (IGRB)  \cite{Ackermann:2014usa}
along the arrival directions of observed UHECRs. 
In this way we attempt to test whether the cumulative signal from yet 
unresolved hard gamma-ray sources 
can be measured.

In contrast to photons, cosmic rays are expected to suffer deflections due to Galactic magnetic field (GMF) 
and extragalactic magnetic fields (EMF) \cite{2013JCAP...01..023F}.  Since the strength and orientation of these magnetic fields are not well known, 
the exact cosmic-ray deflection cannot be well predicted.  
Nonetheless, given such energetic cosmic rays, deflections  are expected to be small,  
in particular for proton primaries.
\rev{In this respect, recent PAO data  \cite{Aab:2014aea} seem to exclude a pure proton composition and instead prefer
a  mixed composition  for the highest energies (protons plus light and intermediate nuclei).
This result is in some tension with TA composition studies \cite{Abbasi:2014sfa} which are instead compatible with 
a pure proton composition, although error bars are larger  in the TA case.
A joint TA and PAO task force has been formed to settle the issue \cite{Abbasi:2015czo}.}
In this work we will assume UHECR to be protons. The analysis is, however, still valid in the case
only a fraction of the observed events are protons. In this case, the rest of non-proton events
would act as a further background for the correlation search and the sensitivity
results reported below would have to be rescaled by the corresponding proton fraction.

The deflection of UHECR in regular and random GMF as well as in the turbulent
EMF can be written as \cite{2013JCAP...01..023F}

\begin{eqnarray}
\delta _{\rm reg, Gal} &=& \left(2.3^\circ \pm 0.24^\circ\right) ~\left(Z/E_{100} \right)\\ \nonumber
\delta _{\rm rand, Gal} &=&  1.3^\circ ~\left(Z/E_{100} \right) \sqrt{\lambda _{rand, Gal}/100 {\rm pc}} \\  \nonumber
\delta _{\rm EMF} &=& 0.15^\circ ~\left( \frac{D}{3.8 {\rm Mpc}} \frac{\lambda _{EMF}}{100 {\rm kpc}} \right) ~\left( B_{\rm EMF}/1 {\rm nG} \right) ~\left( Z/E_{100} \right)
\end{eqnarray}

Here, $Z$ is the charge in units of the proton charge, $E_{100} $ is the energy of the UHECR in units of 100 EeV, 
$\lambda _{rand, Gal}$ is the maximum coherence length of the turbulent Galactic field, 
$D$ is the distance of the extragalactic source from  our Galaxy and $B_{\rm EMF}$ and $\lambda _{EMF}$ are the EMF
mean value and coherence length. 
Given the uncertainty in the MF, from the above equations we take 
as typical expected UHECR proton
deflection  the range $\lsi 4^\circ-6^\circ$.

In addition to actual gamma-ray emission from the sources, 
we should also be sensitive to secondary gamma rays produced from the interaction of UHECRs with the low energy photons which permeate the Universe. There are two important contributions to secondary electromagnetic cascades from UHECR. 
Besides the already mentioned pion production on the CMB,
 the second source of electromagnetic cascades is pair production by protons on low energy photons, $p+\gamma \rightarrow p+ e^++e^-$. 
We consider in particular the second case, 
since the UHECRs survive with minimal energy losses and are accompanied by an electromagnetic cascade. 
In such instances, the additional deflection of the final gamma rays from the source position is quite uncertain 
as it depends on the structure and strength of the EMF \cite{2007Ap&SS.309..465G,2007APh....28..463A}. 
For our search we  assume the same range of angles ($\lsi 4^\circ-6^\circ$ ) as for proton deflections.

The structure of the paper is as follows.  In Section \ref{section:data} we define the {\it Fermi}-LAT and UHECRs datasets to be used along the article.  
In Section \ref{2FHLcorr} we perform the correlation between UHECRs and the 2FHL catalog.  
In sections \ref{ring}-\ref{polspice} we describe the 
search for a possible correlation between UHECRs and diffuse photons using three different analysis methods: 
a) The `ring' method (Section \ref{ring}), in which we use rings around 
the UHECR events 
as our signal region and estimate the background from the surrounding region; 
b)  A search for  possible excesses in the data  over a model of the sky (Section \ref{data-model}); and 
c) A cross-correlation function analysis (Section \ref{polspice})  looking for a signal at different angular scales.  
We end the paper with our conclusions in Section \ref{conclusions}.

\section{Dataset}
\label{section:data}
\subsection{Fermi-LAT}
For the whole-sky {\it Fermi}-LAT data analysis, we 
use 60 months\footnote{The data set covers the time period between 2008 August
4  and 2013 August 4 (239557417 - 397345414 mission elapsed time).} $\rm{P7REP\_CLEAN\_V15}$ events. 
We combine front-converting and back-converting events in a single dataset.
In the parts of analysis where 
we model the sky emission we use the corresponding LAT background model, \texttt{gll\_iem\_v05\_rev1.fit}\footnote{${\rm http://fermi.gsfc.nasa.gov/ssc/data/access/lat/BackgroundModels.html}$} \cite{2016ApJS..223...26A}. 
We also use (see below) the information on the position of sources from the 3FGL catalogue \cite{2015ApJS..218...23A}, which is based on a whole sky analysis of {\rev four years of} P7REP LAT data\footnote{${\rm http://fermi.gsfc.nasa.gov/ssc/data/access/lat/4yr\_catalog/. }$} 
{\rev 
and, in Section \ref{2FHLcorr}, we further use the 2FHL catalog of sources above 50 GeV, which is based on 80 months of Pass 8 LAT data.}

\subsection{Pierre Auger Observatory and Telescope Array}
The UHECR data has been taken from recent PAO \cite{PierreAuger:2014yba} and TA \cite{Abbasi:2014lda} published results.
In order to combine PAO and TA events, 
we have re-scaled PAO events taking into account a $23 \%$
energy scale difference following Ref.~\cite{dmitri}.  
It is worth noticing at this point that work in progress \cite{maris} may yield new modifications to this re-scaling.
{\rev Nonetheless, we have tested explicitly that, for the two actual datasets we use, a rescaling of 23\% is appropriate and brings the TA and PAO
energy spectra in very good agreement.}

For the purposes of this work, we have selected the UHECR data by requiring an energy threshold of $E\gsi60$ EeV in PAO energy scale
or $E\gsi75$ EeV in PAO rescaled energy.

\begin{figure}[ht]
\vspace{-1.0cm}
\begin{center}\includegraphics[width=.85\textwidth]{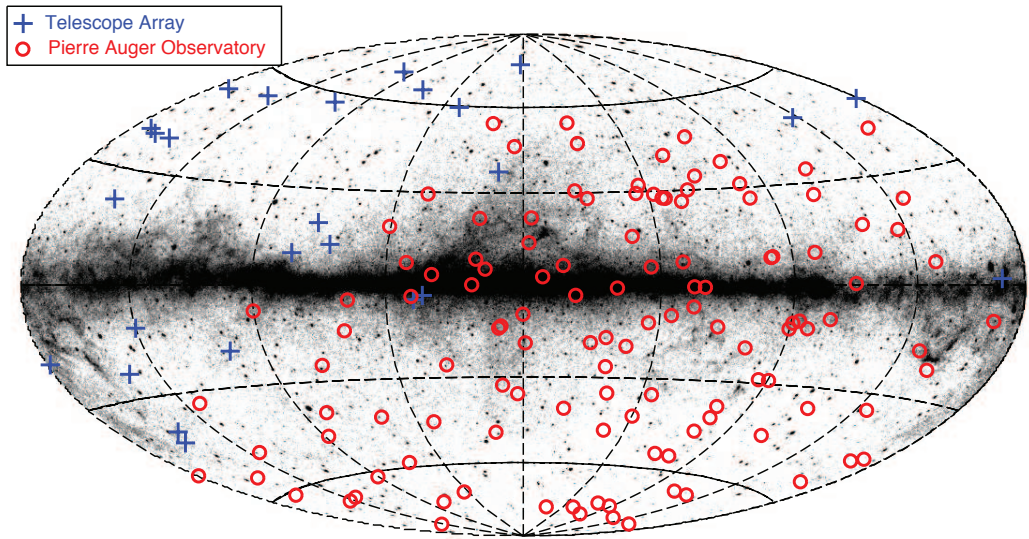}\end{center}
\caption{148 UHECR events with energy greater than 75 EeV in PAO and TA. In the background a map of {\it Fermi}-LAT photons with 
energies above 1 GeV$^a$.}
$^a$ http://fermi.gsfc.nasa.gov/ssc/
\label{uhecrs}
\end{figure}

This yields a total of 148 UHECR directions which are plotted in Fig.~\ref{uhecrs}.  From these directions, 122 come from PAO data and 26 from TA data.   
\rev{We tested also different thresholds at 60 EeV and and 90 EeV and found similar results as for the 75 EeV case. 
Thus, the precise choice of the threshold seems not to be crucial for the analysis.}

\section{Cross-Correlation with the 2FHL} \label{2FHLcorr}

\begin{figure}[t!]
\vspace{-2.0cm}
\centering
\includegraphics[width=.6\textwidth]{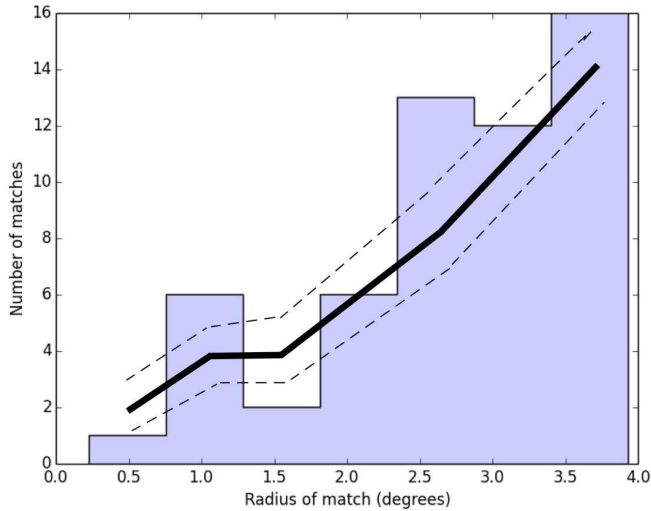}
\vspace{-2.5cm}
\caption{Histogram distribution of nearest angular distance between UHECR directions and 2FHL sources. 
Also shown is the average distribution and 1-$\sigma$ uncertainty band derived from the GRB
sample (thick/dashed lines).}
\label{fig-nestor}
\end{figure}

We compare the UHECR event list with the recently released Second Catalog of Hard {\it Fermi}-LAT sources (2FHL) \cite{Ackermann:2015uya}. The 2FHL lists 360 sources in the 50 GeV--2 TeV range that might be particularly connected to UHECR acceleration. For the cross-matching, we used a similar procedure as the one outlined in  \cite{2010MNRAS.405L..99M}  where  the First {\it Fermi}-LAT Source Catalog (1FGL) \cite{Abdo:2010ru} was used.  Of the 148 UHECRs in our sample, we find 56 events with a 2FHL match within $4^\circ$ (42 coincidences with PAO and 14 with TA). Among the matches, there are 2 events within $4^\circ$ of AP Librae,
which is the only low-frequency-peaked BL Lac  detected 
at energies $E > 100$ GeV \cite{hess}.
However, at a distance of 214 Mpc, AP Librae would be near the edge of the expected GZK horizon.  Figure  \ref{fig-nestor} shows the histogram of the nearest angular distance between a UHECR and 2FHL sources up to  $4^\circ$. 

In order to assess the meaning of this quantity, we can compare it with random samples drawn from directions of arrival of GRBs, which are isotropically distributed over the sky \cite{Meegan:1992xg}. More specifically we build a master GRB file with 2702 GRBs drawn from the BATSE 4B catalog \cite{Preece:1999fv} and 993 GRBs from the Swift BAT\footnote{${\rm http://swift.gsfc.nasa.gov/archive/grb\_table/}$}
(up to UT 2015 October 5) for a total of 3695 events. From this grand list, we draw 500 random samples of 148 events each to compare directly with the 2FHL. 
In order to simulate the events, we use the TA and PAO relative exposures from \cite{Aartsen:2015dml}. Further,  for each sample,  we restrict ourselves to 122 events drawn in the PAO range $-79.2^\circ \leq \,$Dec$\, \leq 42.8^\circ$ and 26 events drawn in the TA range $-1.7 \leq \,$Dec$\, \leq 62.8$, respectively. We find on average  $49 \pm 8$ matches within $4^\circ$ from these random isotropic distributions. No significant
difference is found between actual and random samples 
with matching circles as large as $12^\circ$, indicating no significant 
correlation between UHECRs and the 2FHL.  We also tested for correlation with the subsample of 2FHL sources  with $z<0.1$.
Also in this case no significant correlation was found.
{\rev Our null result in this high-energy band ($> 50$ GeV)  is
  consistent with the lack of significant correlation between UHECRs and 1FGL point sources in the 100 MeV-100 GeV energy range
  \cite{2010MNRAS.405L..99M,jiang2010}. It is important to note, however, that the 1FGL was constructed using the first 11 months of the {\it Fermi} mission, while the 2FHL is based on 80 months of data, and thus includes more sources than the 1FGL in the energy range above 50 GeV.}

\section{Cross Correlation with diffuse gamma-ray emission} 

After having checked for a correlation of UHECRs with resolved gamma-ray point sources,
we here check for a correlation with the gamma-ray \emph{diffuse emission} and thus
with unresolved point sources.
For this purpose, in the remainder of the paper, 
3FGL point sources will always be  
masked using
a $1^\circ$ disk radius around the position of each source.  
An extra complication with respect to the case of correlation with resolved sources is  
the presence of the Galactic diffuse emission which constitutes a (non-isotropic) background
for the correlation with unresolved sources.
In the next section we will employ a methodology where this background is
estimated in a model independent way, directly from the surroundings of each UHECR event.
We will see, however, that this methodology has some limitations.
In Sect.~\ref{data-model}, we thus introduce a second method where the 
Galactic emission is described in terms of  the LAT background model, \texttt{gll\_iem\_v05\_rev1.fit}~\cite{2016ApJS..223...26A}. 
In Sect.~\ref{polspice} we compute the correlation using the formalism of correlation functions.

\subsection{Stacking using a model-independent background estimate} \label{ring}

\subsubsection{Method}

\begin{figure}[t]
\begin{center}\includegraphics[width=.7\textwidth]{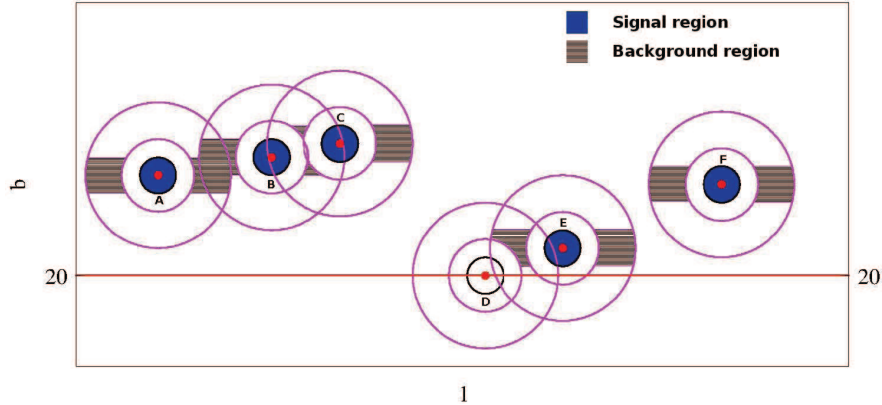}\end{center}
\caption{Illustrative example of signal and background regions definition.   Events A and B have  overlapping  background regions.  
Part of the background regions from events B and C are removed to avoid signal contamination from the closer-than-7$^{\circ}$ UHECR events. 
Event D  has part of its signal region in the Galactic plane and it is excluded from the analysis.  
Event E has an excised background region to avoid signal contamination, even though event D is not taken into account in the analysis.  
Signal and background regions in event F are fully taken into account.}
\label{regions}
\end{figure}

For our stacking, we define the signal region for each UHECR event as a disk around each UHECR direction. 
As reference case we will use a disk radius of 4$^\circ$.  
To estimate the background, we consider an annulus  around each UHECR direction 
which should be close enough to preserve the sky properties, but far enough to avoid signal 
contamination. 
A similar  method was previously used in the analysis of LAT data in the vicinity of dwarf satellite galaxies of the Milky Way \cite{Mazziotta:2012ux} and around low redshift blazars \cite{Chen:2014rsa}.  
We found it suitable to define the background annulus as going from 7$^\circ$ to 12$^\circ$  from its corresponding UHECR event.  
Nonetheless, all calculations in this Section have been repeated using different sets of angular distances and 
we have not found qualitative differences in the final results.  
In particular, we have  tested our method using for signal and background regions the following sets: $0-6^\circ$ and $8^\circ-18^\circ$; $0-8^\circ$ and $10^\circ-20^\circ$; and $0-1.5^\circ$ and $4^\circ-5.5^\circ$; respectively.  
Given the strong gradient in photon density in Galactic latitude  $b$, we select in the background annulus only 
the region which spans the same $b$ as its corresponding signal disk.  
Since it is unlikely that UHECR events could come from the Galaxy, and to reduce Galactic contamination in the gamma-ray emission, 
we  mask regions with  $|b|<20^\circ$.
We also discard any UHECR event if its signal disk is not entirely in the $|b|>20^\circ$ region.  
Finally, as mentioned above, we mask with a 1$^\circ$ radius all sources in the 3FGL catalogue.
Within this definition of signal and background regions for each UHECR direction, 
we still have a complication in the cases where there is overlap between the regions of different UHECR events.  
Therefore, to avoid signal contamination in a background region, 
we remove from the background region all fractions of the annulus that lie closer to 7$^\circ$ from an UHECR direction.  
We plot in Fig.~\ref{regions}, just for visualization purposes, an illustrative distribution of signal disks and background regions 
to show how these regions are defined in a specific case.

The total signal and background regions  are defined as the sum of all the single regions. 
Once those are defined as a function of the UHECR directions, we can compare the number of photons in the signal region 
to the null expectation based on the background region.  
If the photon counts in signal and background regions are $n_s$ and $n_b$, respectively, 
and $c$ is the ratio of solid angle between both regions, then the variable $N_S = n_s - c \, n_b$ 
represents the excess of photons in signal region that would correspond to a true signal.  
Assuming all Poissonian variables, the pull in this variable is
\bea
\frac{ N_S}{\Delta N_S} = \frac{ n_s - c\, n_b }{\sqrt{n_s + c^2\, n_b}} .
\label{pull}
\eea

To test for  the accuracy in the background estimation, we used the model of the Galactic gamma-ray diffuse emission  
provided by the {\it Fermi}-LAT collaboration for P7REP data.
Since, by construction, the background
model does not contain any signal (i.e. a correlation with UHECRs),
$N_S/\Delta N_S$ should be on average compatible with zero.
We found instead that the method
is not unbiased and a systematic shift is present,
increasing at low energies, where more statistics are available.
To take this effect into account, in the following plots,
we enlarge the statistical error band adding to it the systematic shift, as a function of energy,
as derived above.

\label{results}
\begin{figure}[t!]
\centering
\includegraphics[width=.9\textwidth]{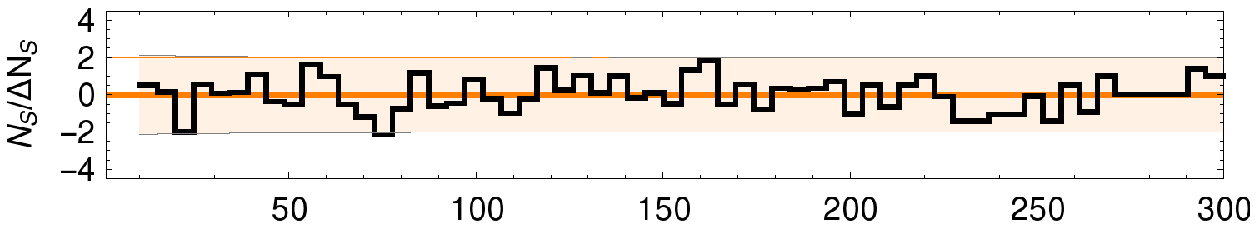}\\
\includegraphics[width=.89\textwidth]{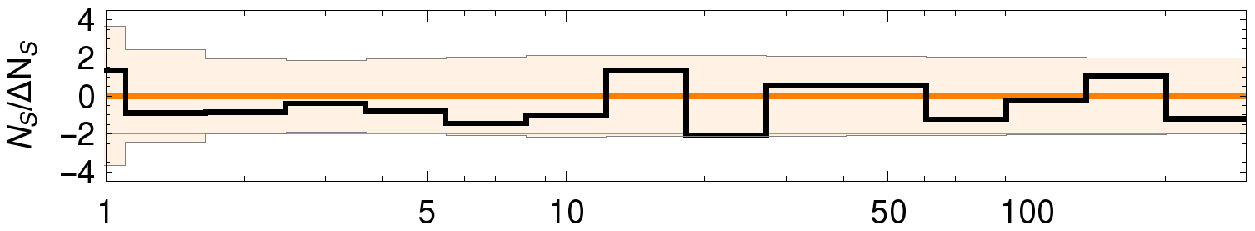}\\
\includegraphics[width=.9\textwidth]{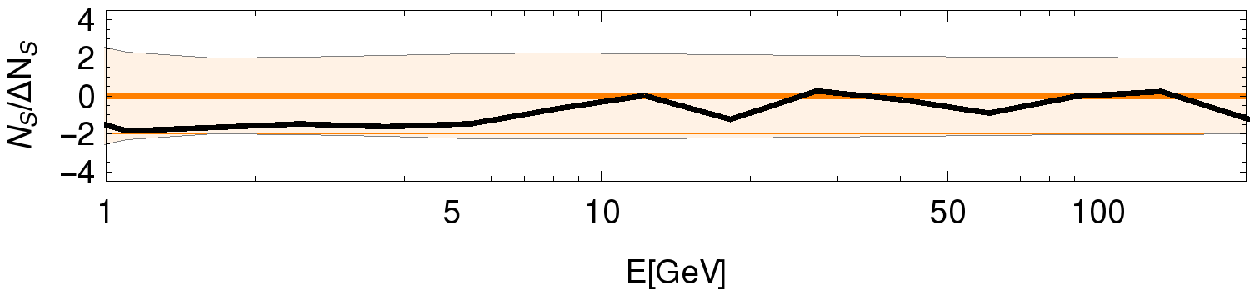}\\
\caption{Pull of the data according to Eq.~\ref{pull} for linear- (upper), Log- (middle) and cumulative-binning (lower).  In the first two plots we compute the pull for each bin, whereas in the third plot we compute the pull for a bin between the given energy and 300 GeV.  In all cases we added, to the $2\sigma$ statistical uncertainty in the pull, the systematic uncertainty band coming from the calculation of the pull in a model with no signal (see text for details).}
\label{main}
\end{figure}

The main reason for this systematic error is the non-uniformity of the
background. 
This has been substantially reduced by selecting as background regions only 
parts of a circle that span the same $b$ as the signal. Nonetheless, although less pronounced, the background
also has gradients in the $\ell$ direction, causing these systematic residual errors. They become evident below
a few GeVs where the statistical uncertainty becomes small due to the large number of gamma-ray events.
In Sect.~\ref{data-model} we present another method which circumvents this issue.

\subsubsection{Results}

Results are shown in Fig.~\ref{main}.
On top of the  2-$\sigma$ statistical error band  we have added the systematic uncertainty described above.
It can be seen that the systematic uncertainty becomes important below a few GeVs,
reaching, in units of the statistical $\sigma$, a value of   $\sim$ 2-$\sigma$ at 1 GeV.
The top panel shows the case of linear binning above  10 GeV,
while the central panel shows the case of a log-energy binning starting from 1 GeV.
These two cases are meant to test the possibility of an excess in a single energy bin, i.e. a \emph{spectral bump} feature.
The bottom panel shows the case  where an excess is searched cumulatively above a given energy threshold,
with the threshold spaced logarithmically in energy.
This last case is meant to test a signal present in many energy bins, as for example the one given by a sources with a power-law  spectrum.
We do not find significant evidence of signal in any case.

\begin{figure}[t!]
\centering
\includegraphics[width=.9\textwidth]{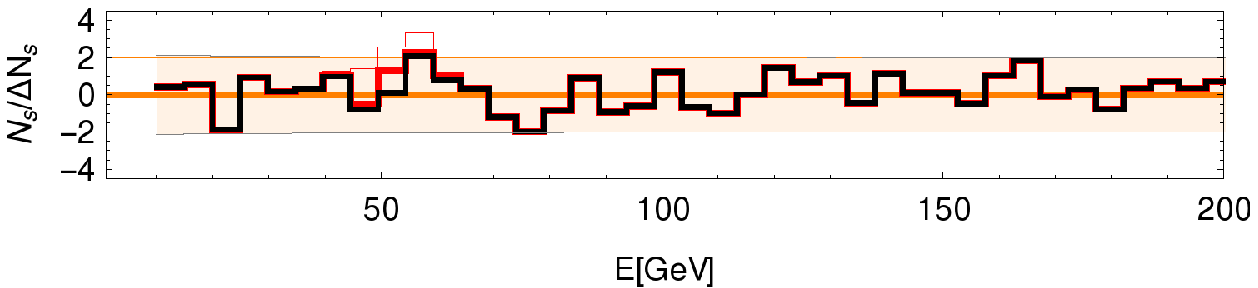}\\
\includegraphics[width=.9\textwidth]{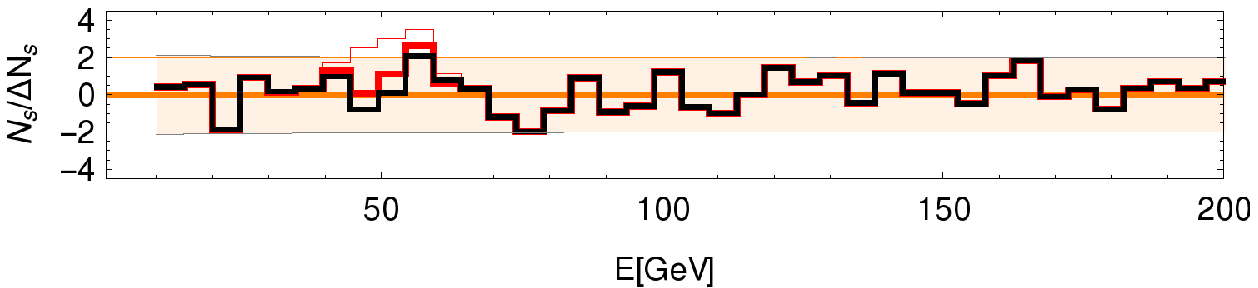}
\caption{Sensitivity of the studied method to a signal localized in the spectrum with $E=50\pm 5$ GeV.  
Upper plot corresponds to the case of 148 sources, and each line from bottom up is data (black), and data plus 100 (red thick) and 300 (red thin) simulated additional photons (see text for details).  
In the lower plot we have assumed 1148 sources, and lines correspond to data (black), and data plus 1k (red thick) and 3k (red thin) photons.}
\label{fake1}
\end{figure}

\begin{figure}[t!]
\centering
\includegraphics[width=.9\textwidth]{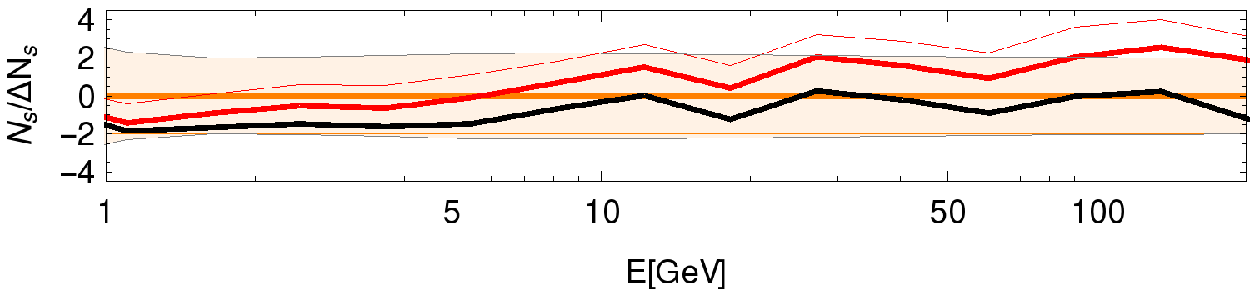}\\
\includegraphics[width=.9\textwidth]{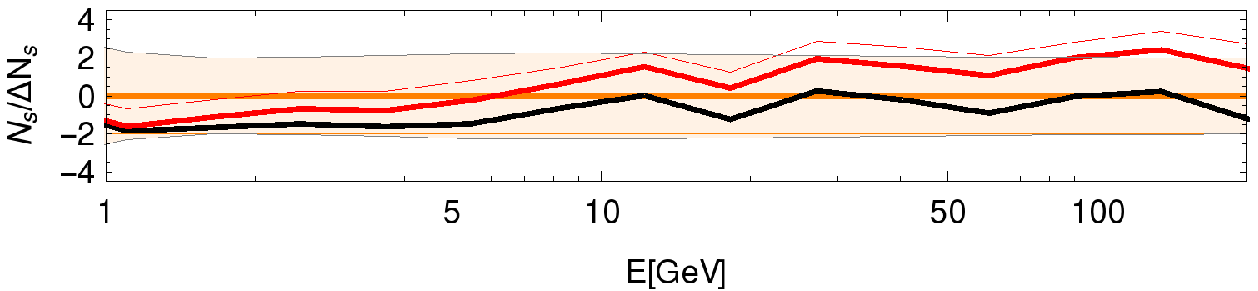}
\caption{Sensitivity of the studied method to a power-law signal with a $E^{-1.7}$ spectrum.   
Upper plot corresponds to the case of 148 sources, and the lines from bottom up are for data (black), 
and data plus 2.5k (red thick) and 5k (red thin) extra photons (see text for details).    
In the lower plot we have assumed 1148 sources, and lines correspond to data (black), and data plus 10k (red thick) and 20k (red thin) extra photons.}
\label{fake2}
\end{figure}

\subsubsection{Flux Sensitivity study}
\label{sensitivity}

In order to study the flux sensitivity of the present method to a hypothetical signal, 
we simulate extra photon events, adding them to the data and re-analyzing the new dataset as a
function of the number of simulated photons.
For each UHECR direction we assign a gamma-ray source drawing its position from a gaussian distribution with sigma  equal to 3$^\circ$
centered on the UHECR. In this way we account for possible deviations due to magnetic fields.
The extra photons are then assigned in the direction of this simulated source.
We consider two setups: i) in the first case we simulate 148 gamma-ray sources, one for each UHECR  
and ii) we simulate an additional 1000 sources randomly located in the sky.
The latter should be representative of a situation in which the UHECR sources are numerous but faint and 
UHECR events have been observed only from a subset of them, while they would still be gamma-ray emitters
close to the detection threshold.
We assume all the gamma-ray sources to be equal and assign the same number of gamma-ray events for each gamma-ray source.
For the energy spectrum of the simulated sources, we studied two cases: i) one of a localized signal in the spectrum (a spectral `line') and ii) one with a power-law spectrum with different spectral indexes.  
In the second case, we consider  the three spectral indexes $2.3$, $2$ and $1.7$ 
in order  to mimick three subclasses of BL Lacs: Low Synchrotron Peaked (LSPs, with an average index of $2.3$), 
Intermediate Synchrotron peaked (ISPs, index $2$) and High Synchrotron Peaked (HSPs, index $1.7$), \cite{2015ApJ...810...14A}.  
The signal localized in the spectrum is a way of explicitly checking our sensitivity in different energy bins, without resorting to a fixed spectral shape. 
For this exercise, we assumed  a Gaussian energy spectrum centered on 50 GeV and with a variance of 5 GeV. 
We generated different samples of extra photons which we added to the real LAT data.  
We plot in Fig.~\ref{fake1} the resulting pull $N_S/\Delta N_S$ for the studied cases.

In Fig.~\ref{fake2} we show the resulting pull in cumulative energy bins for the case in which we simulate sources with an $E^{-1.7}$ power-law spectrum for $E>1$ GeV. 
We have repeated this for softer sources with spectral index $2$ and $2.3$ and found similar results
in terms of number of photons required to see a signal, although the sensitivity can peak at different energies.

We translate these results into flux sensitivity for two scenarios: 
i) in the case of the power-law spectra,
the simulated extra photons are converted into  point-source fluxes, and we compare them with the threshold of different catalogs,
and 
ii) for the case of a spectral line, 
the extra photons are converted into diffuse fluxes (for a uniform region the size of our signal region)
and compared  to the level of the IGRB flux measured in that energy bin. 
This choice corresponds to investigating the physical case of gamma-ray cascades from CR proton primaries. 
To convert counts into fluxes we use a constant (spatially and in energy) average exposure of $1.6 \times 10^{11}$ cm$^2$s, corresponding to the 5-years P7REP\_CLEAN data we use.   We neglect the $\sim 30\%$ variations of the exposure as a function of direction in the sky, and the similar magnitude energy variations in the energy range 1--300 GeV.
We stress that the sensitivity fluxes derived below are only meant as an approximate estimate, with a large uncertainty possibly a factor of a few ($\sim$ 2-3).
We do not attempt to estimate precise sensitivities with a specific confidence level.

\begin{itemize}
\item { \bf Sensitivity to unresolved point sources}

In the case of the continuous spectra from the sources, in Figure \ref{fake2} we see that 5000 extra photons (in the more optimistic case of 148 sources) would produce a detectable ( $\gsi 2\sigma$) signal. This corresponds to a flux sensitivity of $\sim  2 \cdot 10^{-10}$ ph cm$^{-2}$s$^{-1}$ above 1 GeV (obtained by dividing 5k events by the exposure and the number of sources).
This is comparable to the 3FGL\cite{2015ApJS..218...23A} point-source sensitivity above 1 GeV.
In terms of the flux sensitivity above 10 GeV this translates to $\sim 4 \cdot 10^{-11}$ ph cm$^{-2}$s$^{-1}$ for a spectral index of $1.7$, 
\rev{comparable to the  1FHL point-source sensitivity of $\sim 5 \cdot 10^{-11}$ ph cm$^{-2}$s$^{-1}$ above 10 GeV.} Above 100 GeV the sensitivity of the analysis translates to $\sim 3 \cdot10^{-12}$ ph cm$^{-2}$s$^{-1}$ for a spectral index of $1.7$ , to be compared with the 1FHL  point-source sensitivity of  $\sim 3 \cdot 10^{-11}$ ph cm$^{-2}$s$^{-1}$, and the 2FHL one which is a factor 2-3 better than the 1FHL.
We see that, as expected, for hard sources (index 1.7) and high energy threshold searches our method is more sensitive than the catalog searches. 
In a more realistic case of 1148 simulated sources, the point-source sensitivity changes as $\sim 10^{-10}$ ph cm$^{-2}$s$^{-1}$ above 1 GeV,
which is within a factor of 2 from the previous case, and thus compatible within our quoted uncertainties of a factor 2-3.

\item {\bf Cascade photons}

In this case we take into account the size of our signal region, which is 0.97sr considering all the UHECR events. 
With 300 extra events in the $50\pm 5$ GeV energy bin (which from Fig.~\ref{fake1} are approximately necessary to give a $\gsi 2\sigma$ signal with our method) within this area and with the above exposure, the resulting flux sensitivity is $\sim 2 \cdot 10^{-9}$ ph cm$^{-2}$s$^{-1}$sr$^{-1}$. The IGRB flux in the energy bin 36-51 GeV is $1.1\pm0.1 \cdot 10^{-9}$ ph cm$^{-2}$s$^{-1}$sr$^{-1}$ \cite{Ackermann:2014usa}, roughly comparable in value to our estimated sensitivity.   
Since, by different methods (see for example \cite{Ajello:2015mfa}), the unexplained fraction of the IGRB
 is already constrained to be below $\sim$ 10-20\%, we conclude that our method 
is not very sensitive to diffuse emission from UHECRs.
This is likely a consequence of the method itself,  which seems
better suited for point-source searches, rather than for purely diffuse searches.
Indeed, in the globally stacked data from the signal region, the isotropic emission is quite subdominant
with respect to the Galactic emission and the sensitivity is poor.

\end{itemize}

\rev{Our search is complementary to other studies which investigated the connection between UHECRs and 
gamma rays using  the {\it intensity energy spectrum} of the IGRB. 
In particular, constraints on the contribution of population of UHECR sources to the diffuse gamma-ray signal can be obtained by requiring that the cumulative cascade photon flux summed over all the UHECR sources does not exceed the IGRB flux (see \cite{Ahlers:2011sd,2011PhLB..695...13B,Berezinsky:2016jys}). 
This constraints could become even stronger if only the fraction of the IGRB which is of genuinely diffuse origin -- as opposed to being consistent with unresolved point source contribution, is used as an upper limit. For example in \cite{TheFermi-LAT:2015ykq} it was found that diffuse contribution can be as low as 14\% above 50 GeV. 
These constraints disfavour, in the pure proton scenario, UHECR sources evolving as the star formation rate or as GRBs, 
while non evolving sources such as BL Lacs are favoured \cite{Gavish:2016tfl}.
In the future, it might be interesting to perform a more accurate joint study of the two constraints.}

\begin{figure}[t!]
\centering
\includegraphics[width=.48\textwidth]{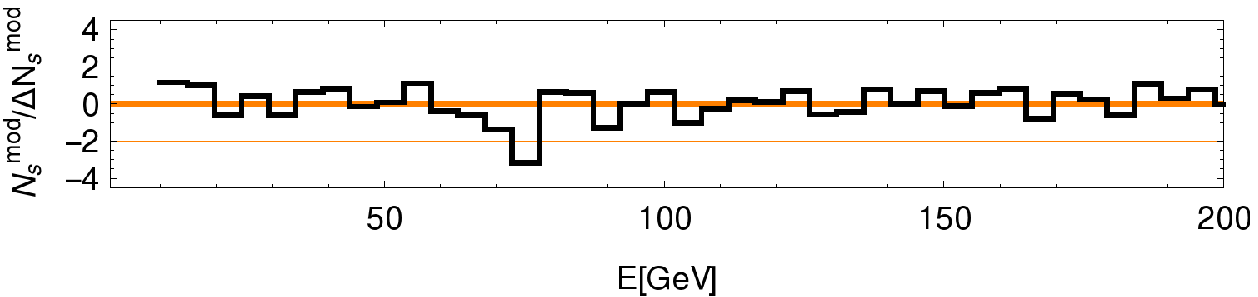}\includegraphics[width=.48\textwidth]{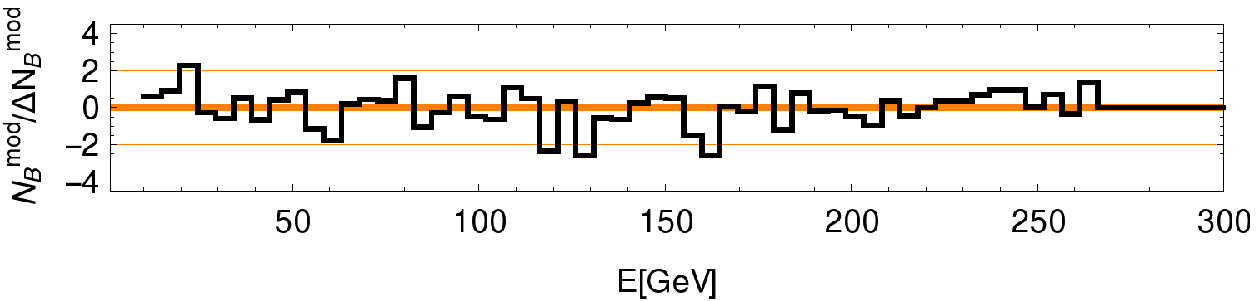}\\ 
\includegraphics[width=.48\textwidth]{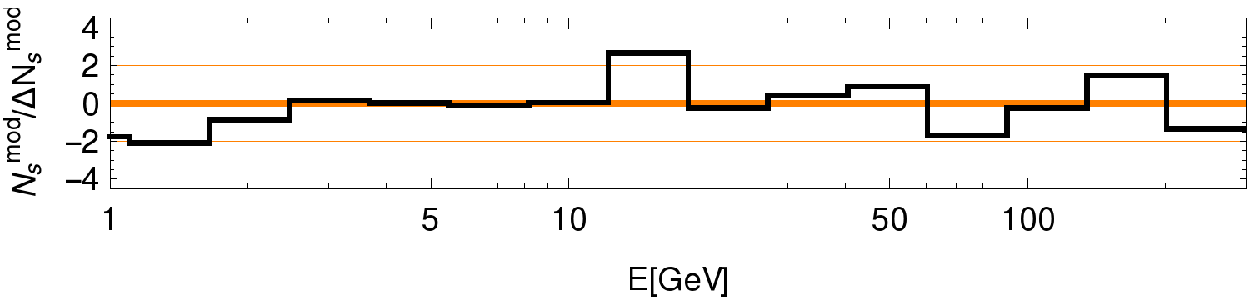}\includegraphics[width=.48\textwidth]{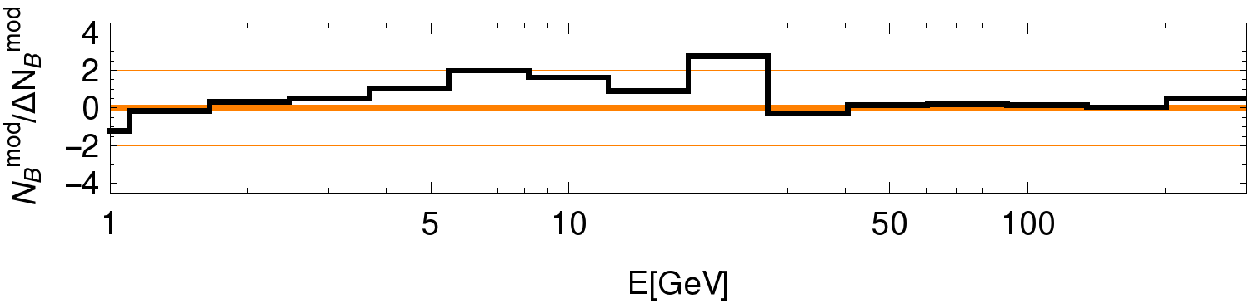}\\ 
\includegraphics[width=.48\textwidth]{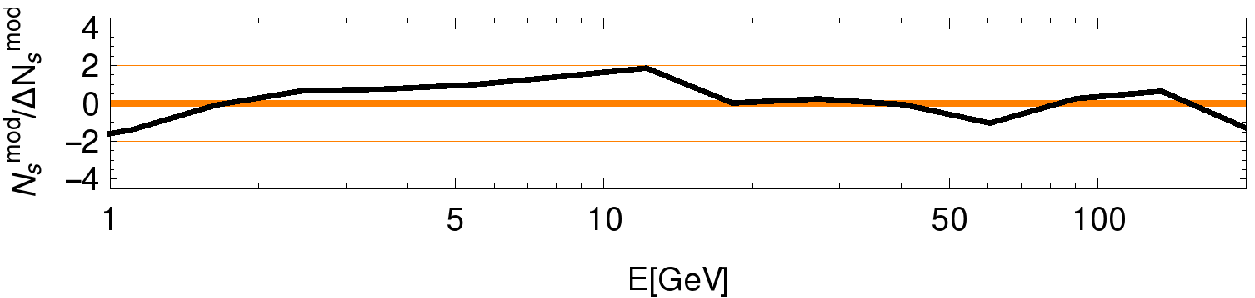}\includegraphics[width=.48\textwidth]{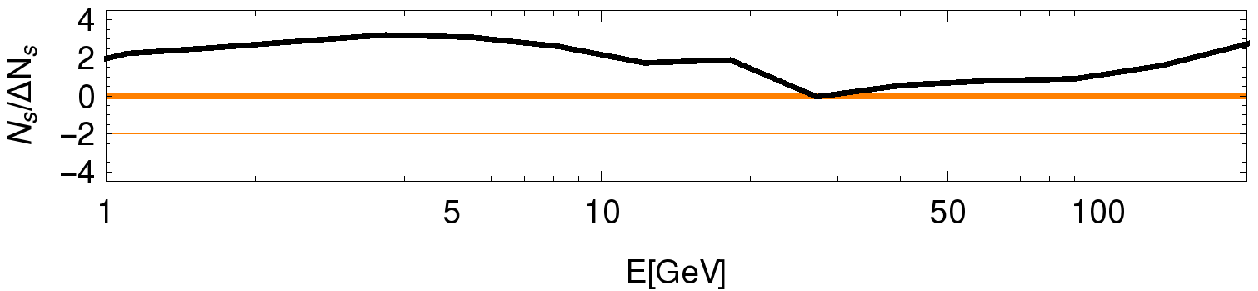}\\ 
\caption{Data - model pull according to Eq.~\ref{datamodel}, where no uncertainty is assigned to the model.  Left plots correspond to signal region, and right plots to background region in order to validate an eventual deviation in signal region.  We have plotted different binning as in Fig.~\ref{main} in each row.}
\label{datamodelfig}
\end{figure}

\subsubsection{Cen A and the TA Hotspot}
\label{section:special}

\begin{figure}[t!]
\centering
\includegraphics[width=.9\textwidth]{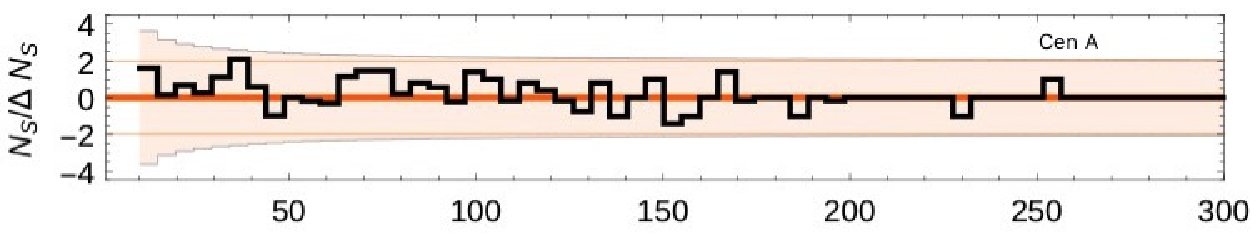}\\
\includegraphics[width=.9\textwidth]{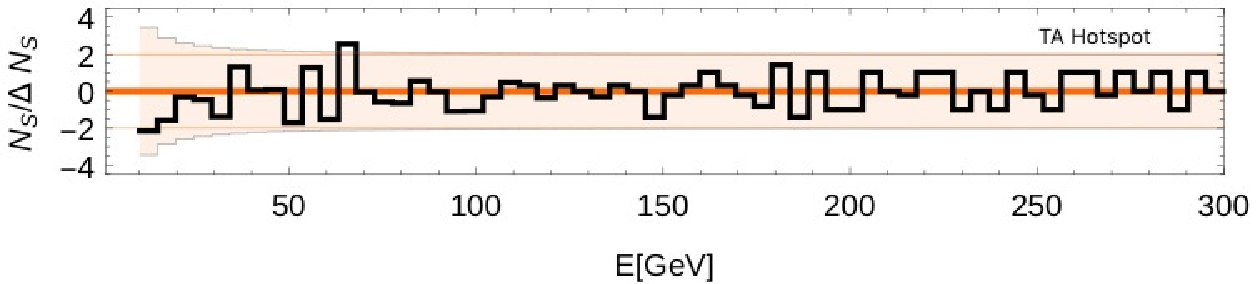}
\caption{Pull of the circular regions centered on Cen A and the TA hotspot
(upper and lower plots respectively).
}
\label{special}
\end{figure}

{\rev As a complement to the all-sky analysis,
  we also search for excesses of unresolved gamma-ray 
  emission around the Cen A and TA hotspot regions.  The nearby Cen A radio
  galaxy is a leading candidate for UHECR acceleration and has shown tentative
  evidence for an excess of PAO events within 18$^\circ$ \cite{abreu2010}. Additionally,
  the Telescope Array has observed a statistically significant
  hotspot of events in a region which stretches at least 20$^\circ$ in radius
  in the northern hemisphere \cite{Abbasi:2014lda}.

Given the possibility that these regions could be the main sources of
UHECRs, we perform two separate analyses using the same method as
before but in this case using {\it only one} direction in the sky.
For Cen A, we consider a signal region of disk radius 9$^\circ$, and a background region as an annulus going from 12$^\circ$ to 17$^\circ$, centered on
coordinates $(\ell,b)=(309.5^\circ,19.4^\circ)$. We chose this 
signal region in order to minimize contamination from diffuse Galactic plane
emission below $|b|=10^\circ$. 
As Fig.~\ref{special} shows,
the result is within the expected uncertainty for this setup, suggesting no
significant gamma-ray enhancement.

Note that in this search the core of Cen A is a masked with a disc of 1$^\circ$  radius similarly to the other 3FGL sources. 
Cen A also presents diffuse gamma-ray emission from its lobes \cite{Fermi-LAT:2010llz,2016A&A...595A..29S}, 
which  extends approximately 10$^\circ$ along the north-south direction. 
In principle this emission is a background for our analysis and it could be modeled with an appropriate template. 
Nonetheless, we see no excess even though we leave this component unmasked and unmodeled. 
This is partly due to the fact that the lobes contribution is very weak when diluted in our large signal region, 
and to the fact that our method is intrinsically not very sensitive to diffuse signals, as discussed in the previous section.

For the TA hotspot, we consider a signal region
centered in $(\ell,b)=(177^\circ,50^\circ)$ with a 20$^\circ$ radius and a background annulus stretching from 25$^\circ$ to 35$^\circ$.  We show the results for this 1-direction analysis in Fig.~\ref{special}.  Again, no significant excess is found in
our analyses.

We also repeated the search for both Cen A and the TA hotspot using the background model method described in the next section, 
again finding no significant correlation.}

\subsection{Stacking using a model of the background}\label{data-model}

As an alternative measure of a possible excess in the signal region, we have also computed the difference between the number of events in the signal region and the number of events predicted in the same region by the {\it Fermi}-LAT model of the sky ($n_s^{mod}$).
To this purpose, 
we  used the model of the Galactic gamma-ray diffuse emission  
provided by the {\it Fermi}-LAT collaboration for P7REP data,  \texttt{gll\_iem\_v05\_rev1.fit},
and the model of isotropic emission \texttt{iso\_clean\_v05\_rev1.txt}, and 
we convolved them with the instrument response functions to obtain maps of expected counts as function of energy  ($n_s^{mod}$).
Analyzing residuals with respect to an assumed background  model, we expect that the gradient effects described for the previous method
should be significantly suppressed, together with the corresponding bias below $\sim$ 5 GeV. 
In this case, we have defined the pull in the variable $N_s^{mod}= n_s - n_s^{mod}$ as
\bea
\frac{N_S^{mod}}{\Delta N_S^{mod}} = \frac{n_s - n_s^{mod}}{\sqrt{n_s}}.
\label{datamodel}
\eea
Note that this formula differs from the one of Eq.~\ref{pull}, since now only the counts in the signal regions
are involved.
In order to validate a possible excess in this variable as a true signal, 
we have also defined a similar variable but for the background region, 
where we do not expect an excess coming from a signal.  
We have plotted in Fig.~\ref{datamodelfig} the outcome of these variables and found a fairly good agreement between data and model.  
As expected, in this case the systematic effect described for the model-independent background stacking is not present.
We have thus  plotted only  the $\pm 2 \sigma$ statistical uncertainty band coming from the data.
Note that in the low energy bins, which are the ones with the largest statistics, a percent level systematic bias in the model would generate a shift in the pull up to 2.
Systematics, indeed, might still be present due to the model not perfectly describing the Galactic background.
Nonetheless, since we did not find a significant signal with this method, we didn't investigate further possible residual systematic model uncertainties.

\begin{figure}[t!]
\centering
\includegraphics[width=.48\textwidth]{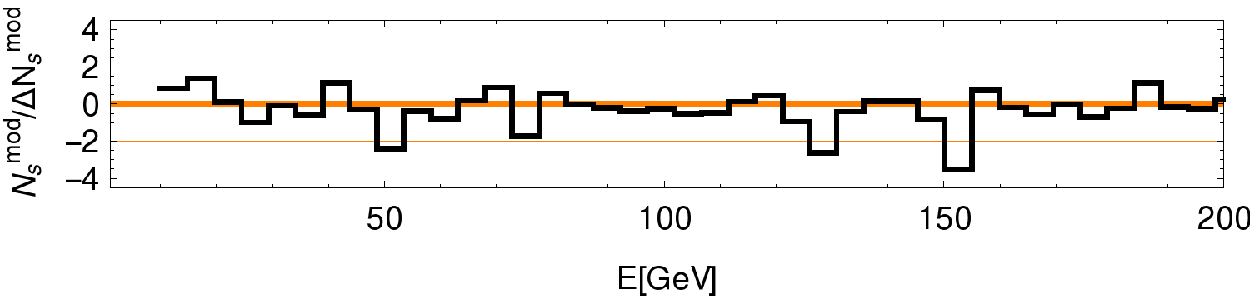}\includegraphics[width=.48\textwidth]{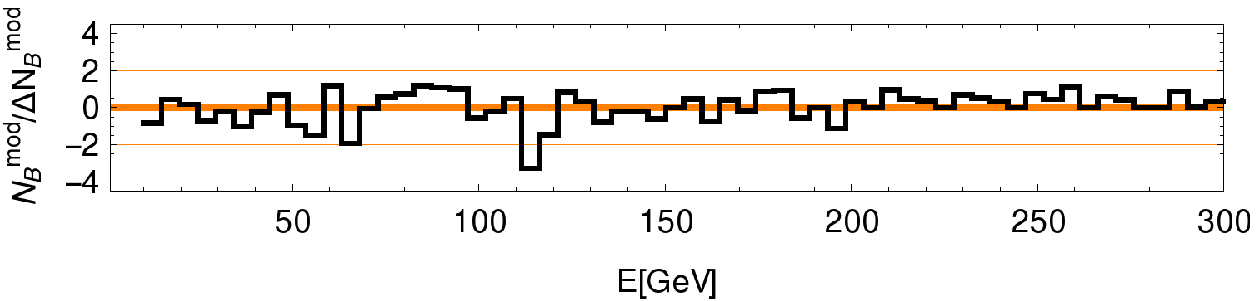}\\ 
\includegraphics[width=.48\textwidth]{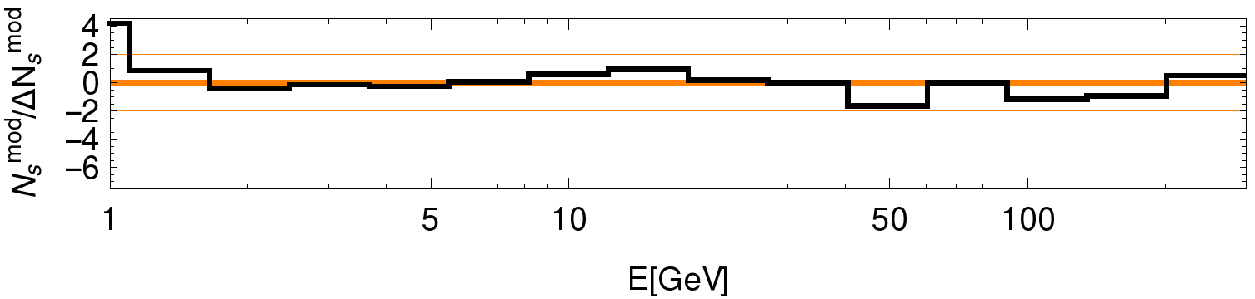}\includegraphics[width=.48\textwidth]{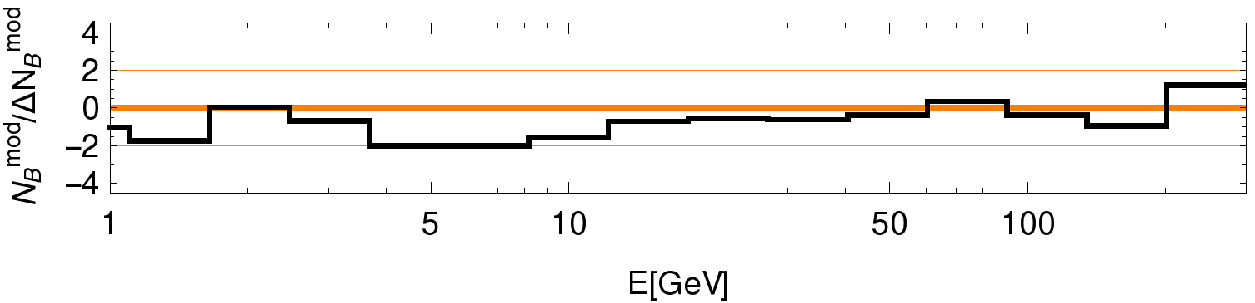}\\ 
\includegraphics[width=.48\textwidth]{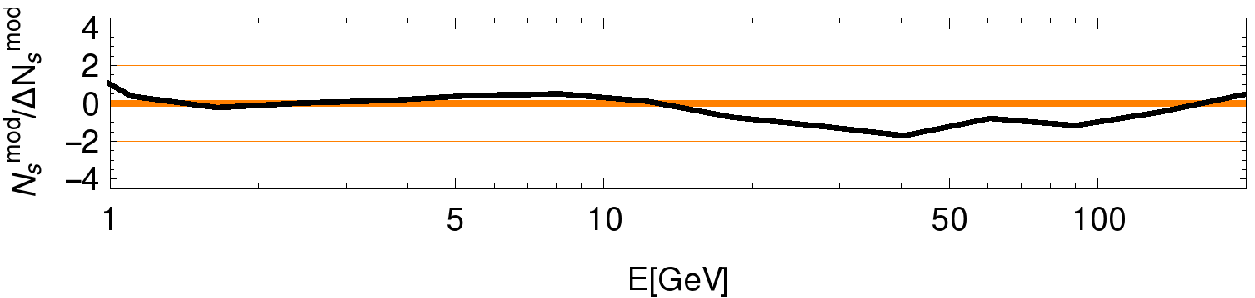}\includegraphics[width=.48\textwidth]{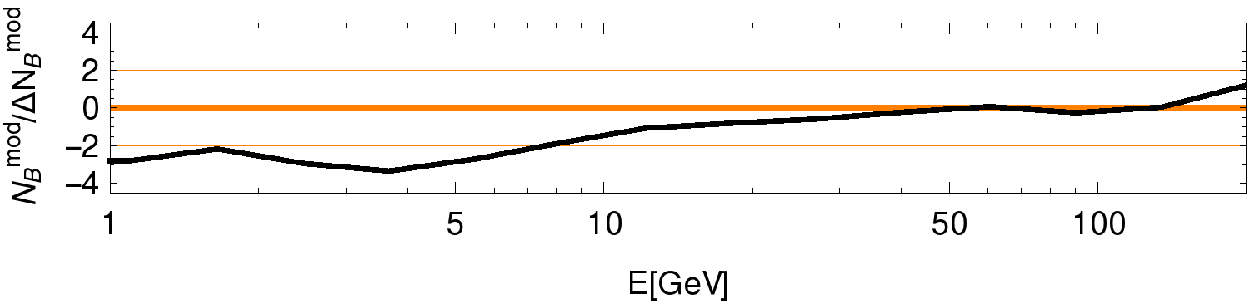}\\ 
\caption{Same plots as in Fig.~\ref{datamodelfig}, but replacing data by a Monte Carlo simulation of the sky. We see that even in the absence of a signal one could expect deviations in the low energy bins.}
\label{mcfig}
\end{figure}

Finally, to further explore the validity of this method, we have repeated the same method but replacing the data 
 by a Monte Carlo simulation of the sky realized with \texttt{gtobssim} based on the same model of the sky emission.  
Point sources are also included in the simulation, and they are masked similarly as in the real data case.
 We have found similar results, including deviations at low energies.  We show the outcome of this test in Fig.~\ref{mcfig}.

\subsubsection{Sensitivity study}
\begin{figure}[t!]
\centering
\includegraphics[width=.48\textwidth]{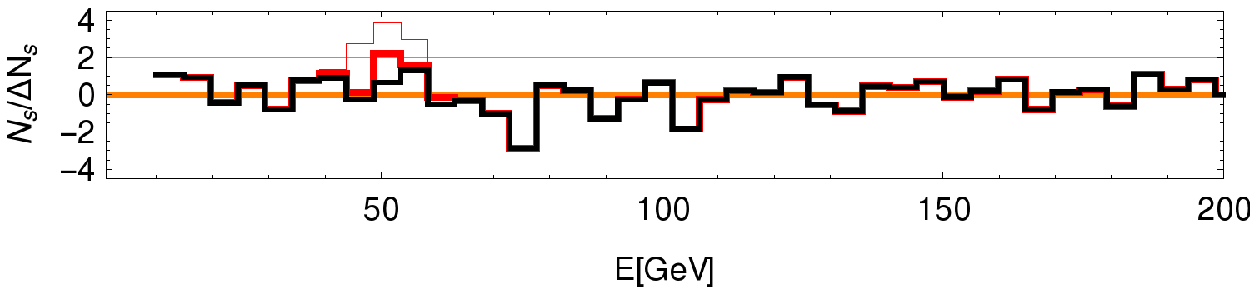}\includegraphics[width=.48\textwidth]{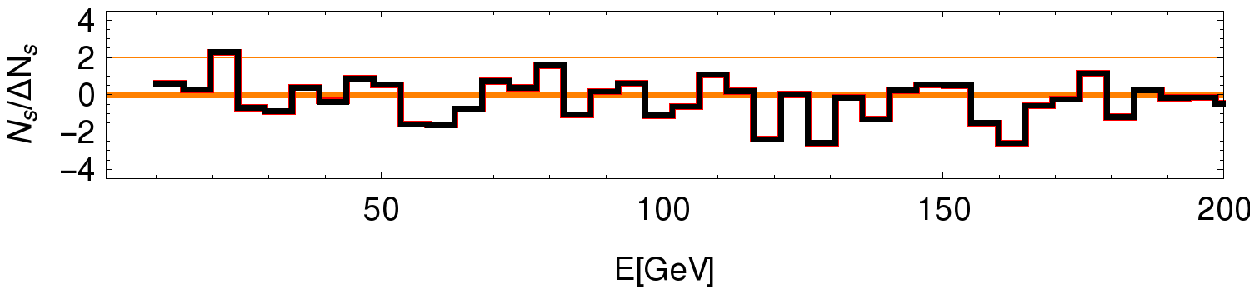}\\ 
\includegraphics[width=.48\textwidth]{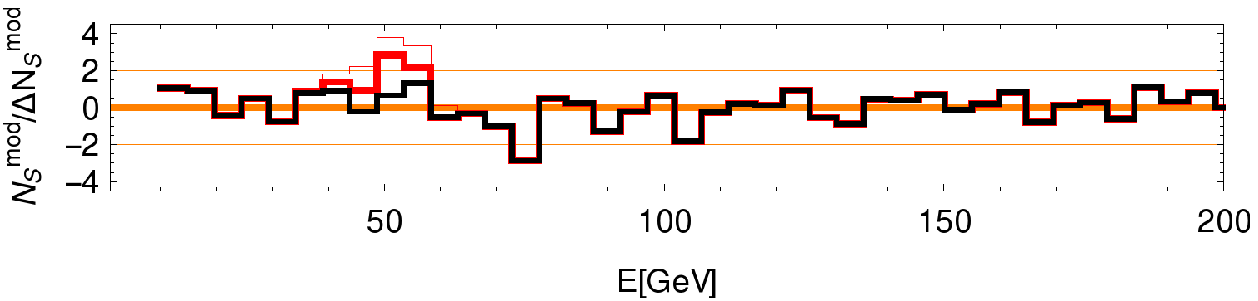}\includegraphics[width=.48\textwidth]{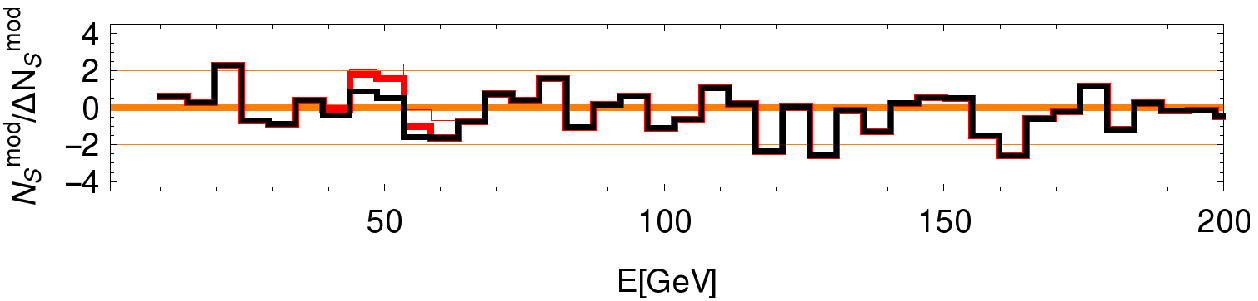}\\ 
\includegraphics[width=.48\textwidth]{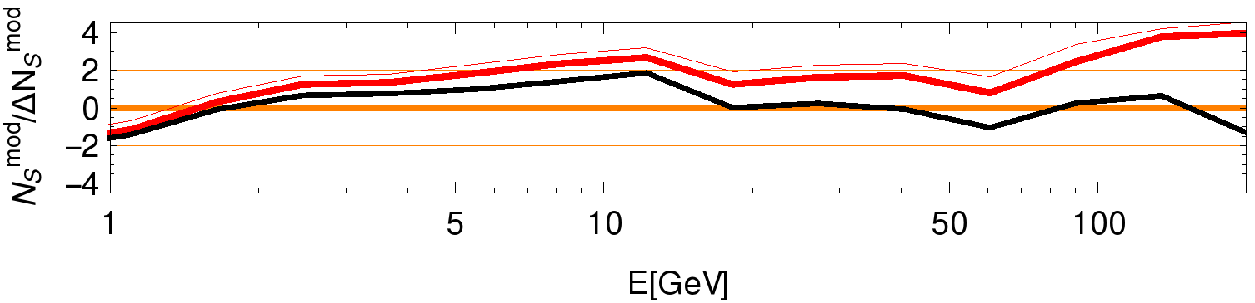}\includegraphics[width=.48\textwidth]{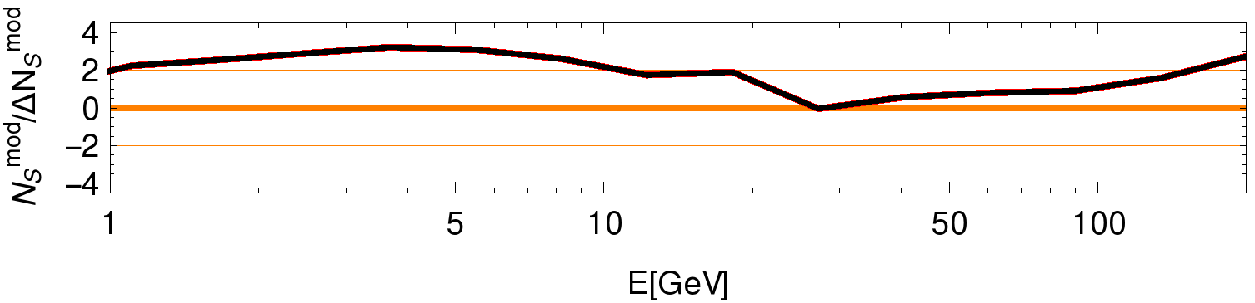}\\ 
\includegraphics[width=.48\textwidth]{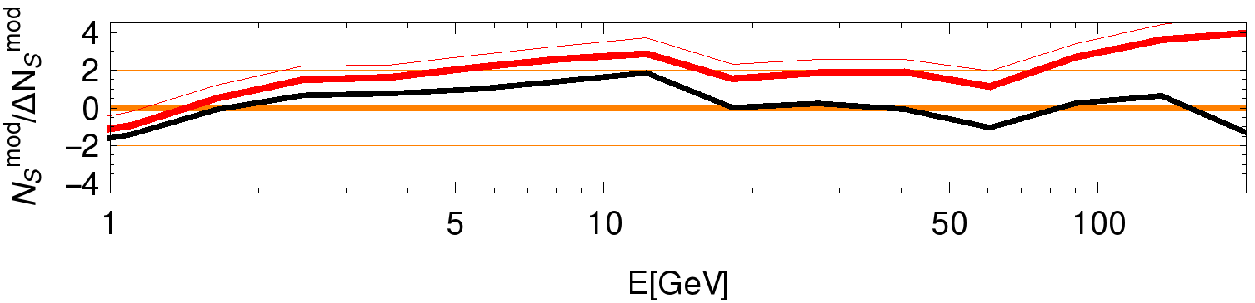}\includegraphics[width=.48\textwidth]{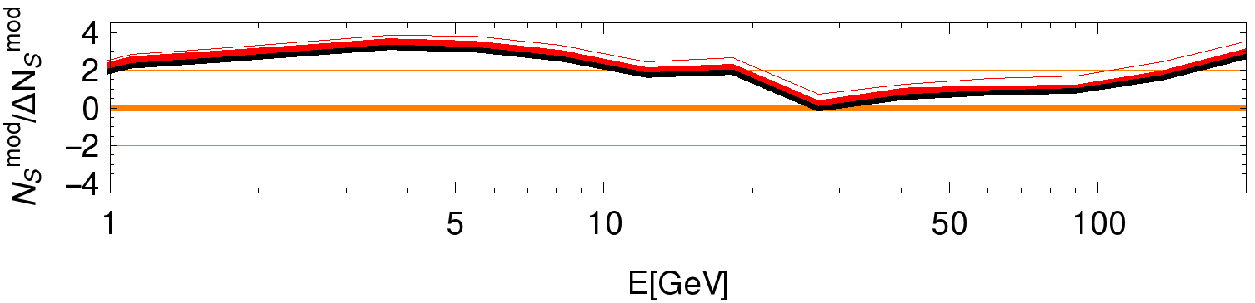}\\ 
\caption{Sensitivity to a simulated signal in the (data-model)/$\sqrt{\mbox{data}}$ method described in Sect.~\ref{data-model}.  
Left plots correspond to signal-only regions and right plots to background-only regions.  
Upper 2 rows correspond to a signal  localized in energy at 50$\pm$5 GeV and differential energy binning, 
and bottom 2 rows to a power-law signal  $\propto E^{-1.7}$ and cumulative energy binning.  
Rows 1 and 3 correspond to 148 UHECR sources, whereas rows 2 and 4 to 1148 sources (1000 extra sources), similar to Sect.~\ref{sensitivity}.  
Red-thick and red-thin lines correspond to different number of extra photons, whereas black is for the original data.  
In row 1 the red-thick line corresponds to 100 extra photons and the red-thin line to 300 extra photons; in rows 2 and 3 to 500 and 1k, respectively; 
and in the bottom row to 2.5k and 5k, respectively.  
Note how, as  expected, no excess appears in background plots in rows 1 and 3. 
For the case of extra UHECR sources (right panels in rows 2 and 4) the pull is sensitive to some signal since many of these extra sources
 fall in the background region (which is always defined only using the original 148 UHECR directions).
Nonetheless, the signal is not  seen at a significant level in this case.}
\label{fake3}
\end{figure}

In Fig.~\ref{fake3} we show the numbers of extra photon events needed to reach the signal detection in the stacking method which relies on a LAT background model\footnote{It has to be noted however that in the case of the signal detection in this method, systematic uncertainties would need to be re-examined.}. 

For the case of the simulated power law signal we see that this method is more sensitive by about a factor of five (i.e.  $\sim$1000 events are needed for a $\gsi 2 \sigma$ detection in this method, as apposed to 5000 events with the previous method, for the case of 148 signal directions). This translates to a flux sensitivity above 10 GeV of $\sim 8 \cdot 10^{-12}$ ph cm$^{-2}$s$^{-1}$ for a spectral index of 1.7 (to be compared to the \rev{$5~\sigma$} 1FHL sensitivity of $\sim 10^{-10}$ ph cm$^{-2}$s$^{-1}$).   

In the case of the simulated spectral line signal we observe only a minor  improvement in this method, compared to the sensitivity calculated in Section \ref{sensitivity}.

\subsection{Cross-correlation analysis} \label{polspice}

Beside the method described above, we also checked for a correlation between UHECRs and the IGRB using the standard formalism of correlation functions.
Both correlation in real space (cross-correlation function CCF) and in harmonic space (correlation angular power spectrum CAPS) have been considered.
For simplicity we have considered two extreme choices of energy ranges to explore, namely $E>1$ GeV and $E>50$ GeV.
The two gamma-ray maps have been `cleaned' from the Galactic foregrounds using explicitly the Galactic foreground model.
Details of the cleaning procedure are as described in the  work  Xia et. al \cite{Xia:2015wka}.
To calculate the correlation and, most importantly, the error bars of the derived CAPS and CCF we use
the software Polspice\footnote{${\rm http://www2.iap.fr/users/hivon/software/PolSpice/}$}. 
Again, the analysis details are the same as  \cite{Xia:2015wka}, to which we refer the reader for a full description
of the technical procedure. 
To derive the correlation we mask the Galactic Plane using different Galactic latitude cuts, of 20$^\circ$, 30$^\circ$, 40$^\circ$, 
and we mask 3FGL point sources
with a disc of 1$^\circ$ radius.

\begin{figure}[t!]
\centering
\includegraphics[width=.48\textwidth]{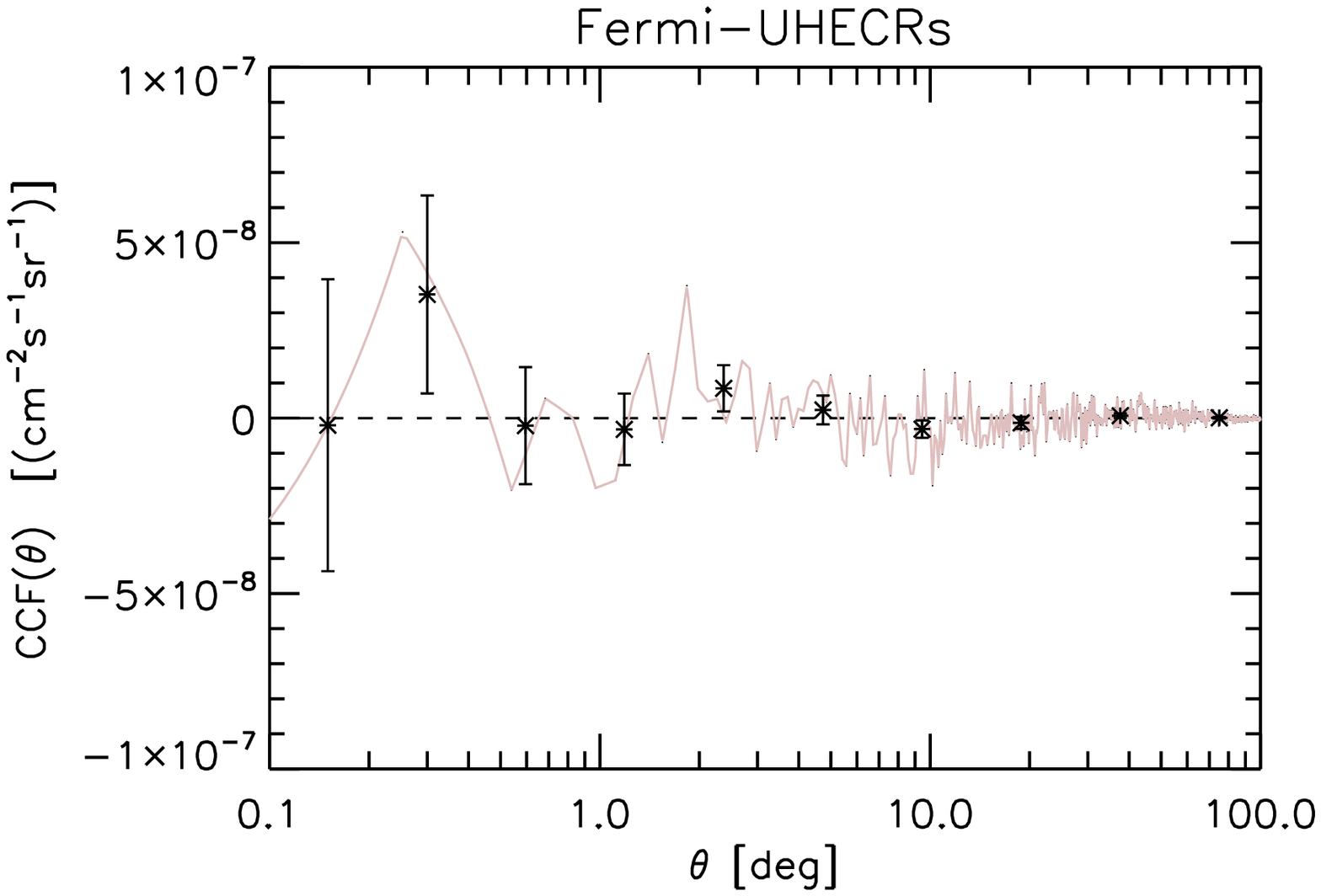}
\includegraphics[width=.48\textwidth]{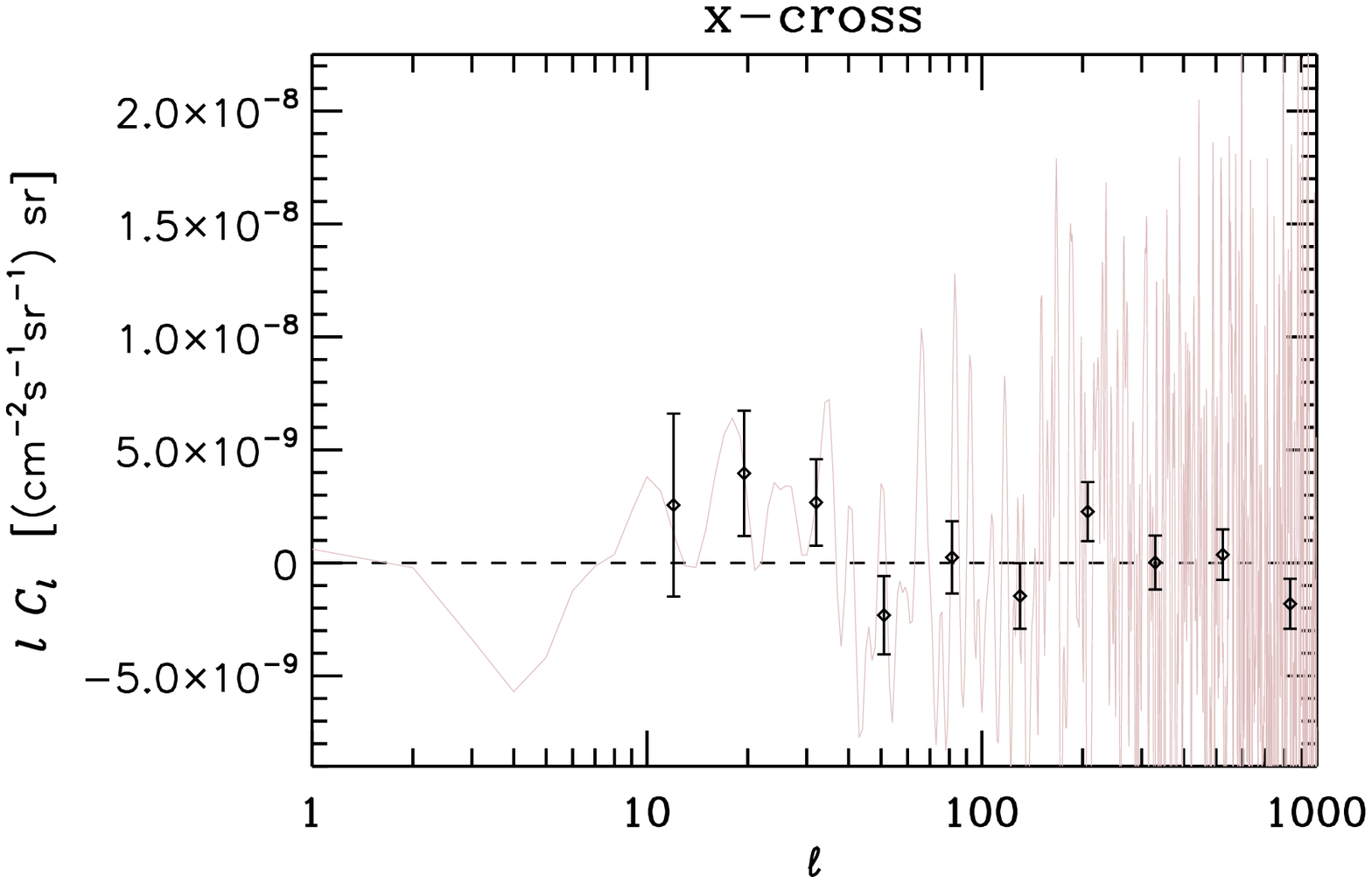}
\\ 
\includegraphics[width=.48\textwidth]{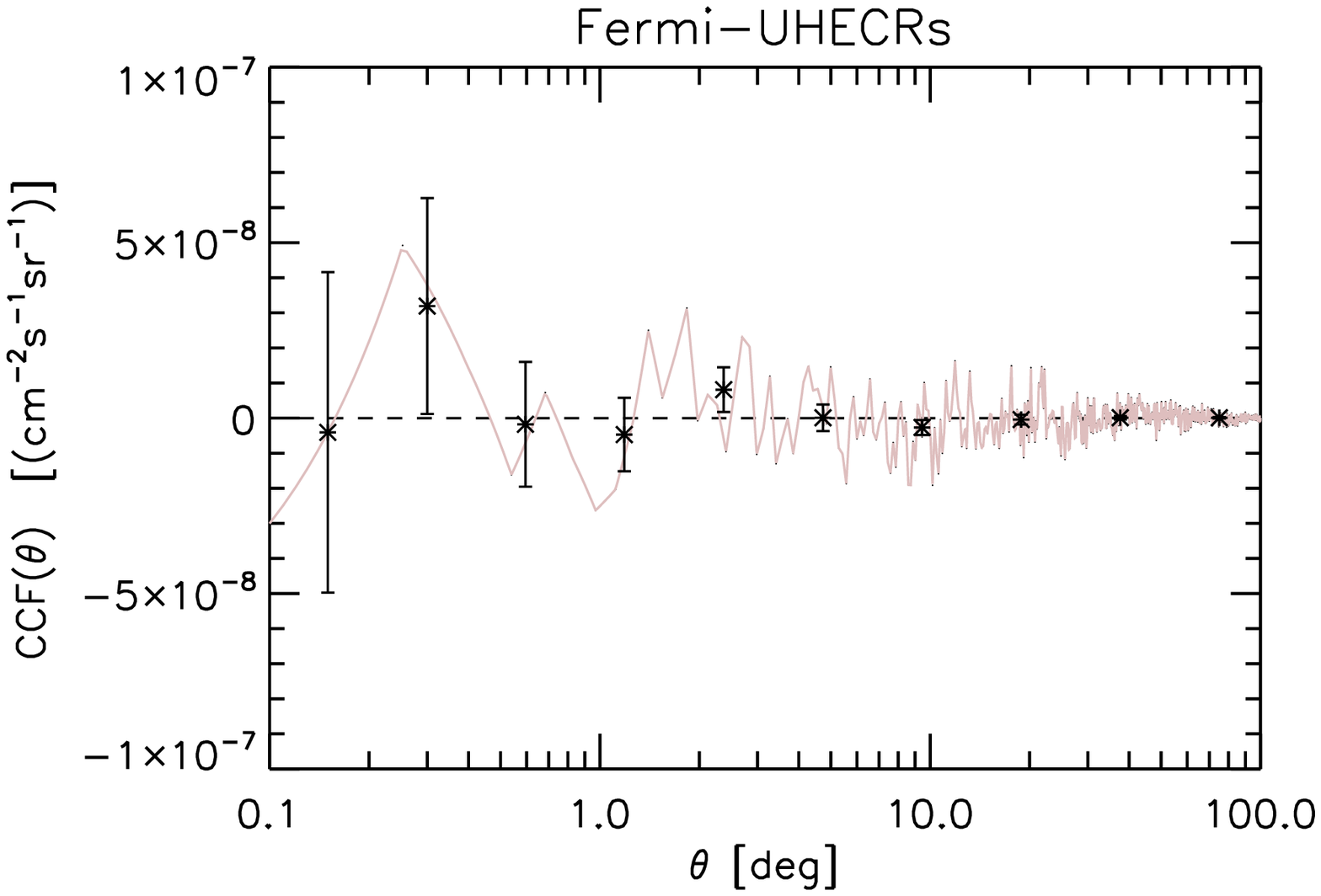}
\includegraphics[width=.48\textwidth]{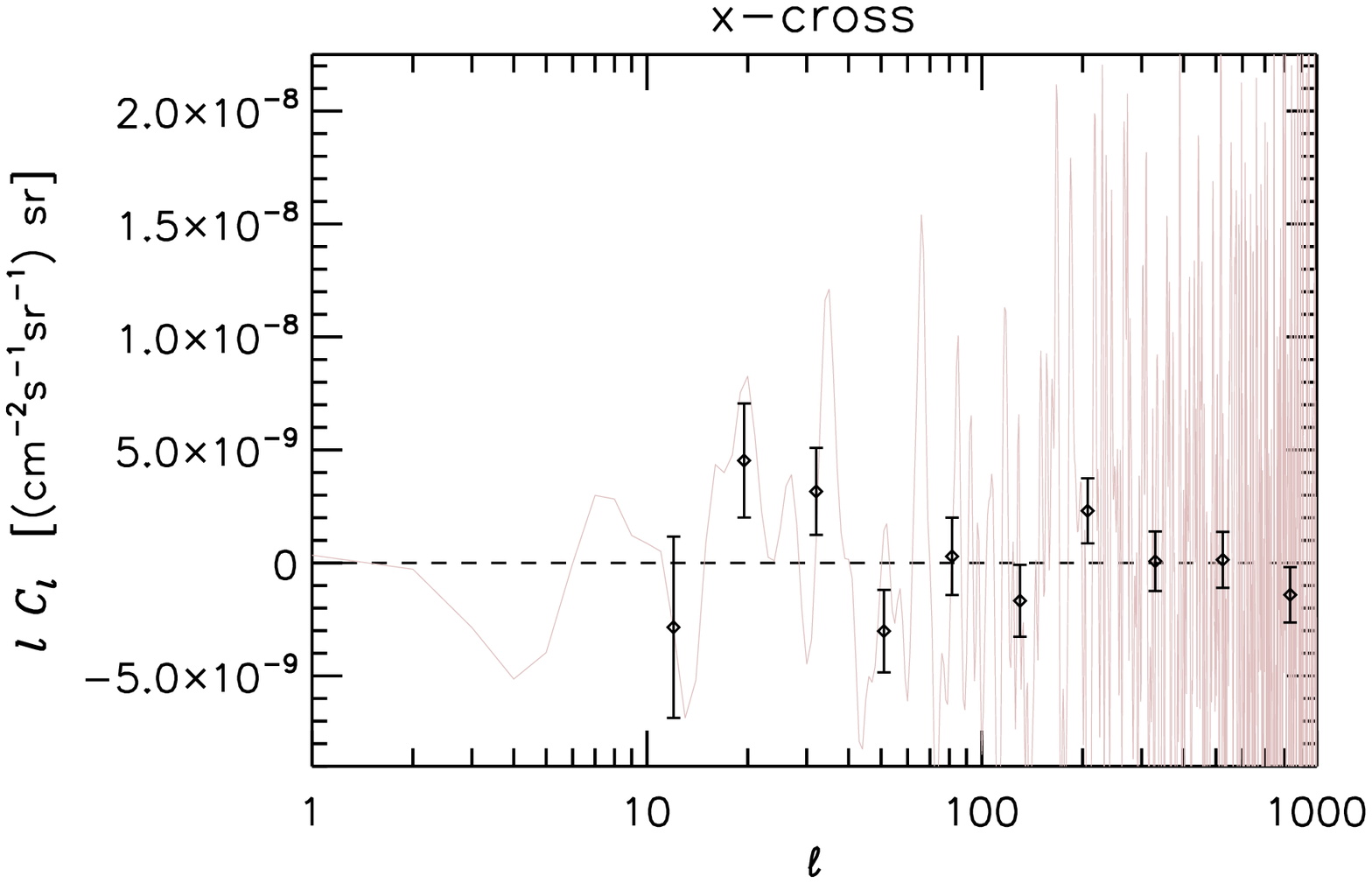}
\caption{Cross-correlation functions (left column) and correlation angular power spectra (right column) between 
the diffuse gamma-ray emission for $E>1$ GeV and UHECR directions.
Upper panels are for a Galactic latitude cut of $|b|>20^\circ$ and lower panels for $|b|>30^\circ$.
In all panels the thin colored continuous line shows the unbinned angular spectra or correlation functions.}
\label{polspice1}
\end{figure}

\begin{figure}[t!]
\centering
\includegraphics[width=.48\textwidth]{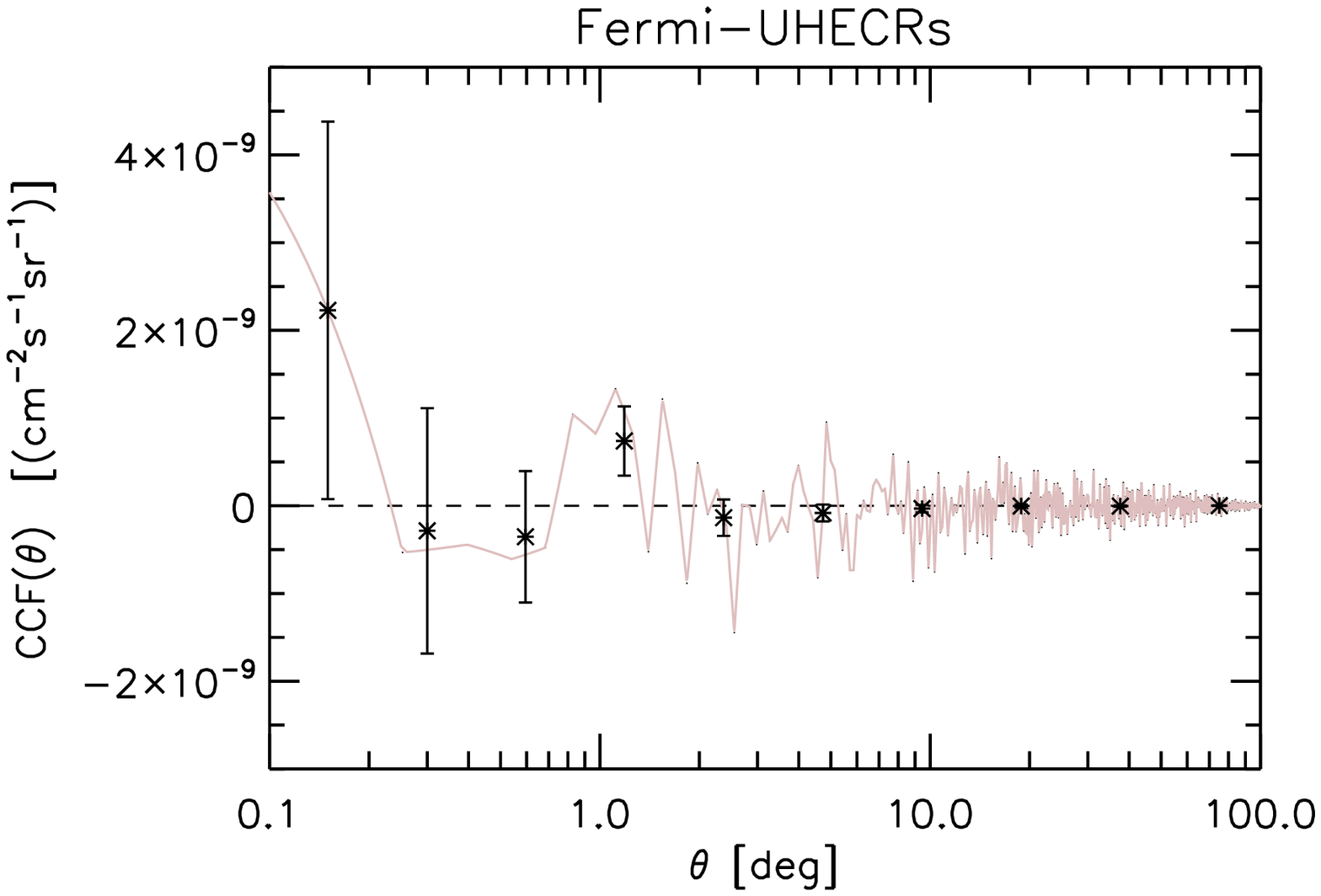}
\includegraphics[width=.48\textwidth]{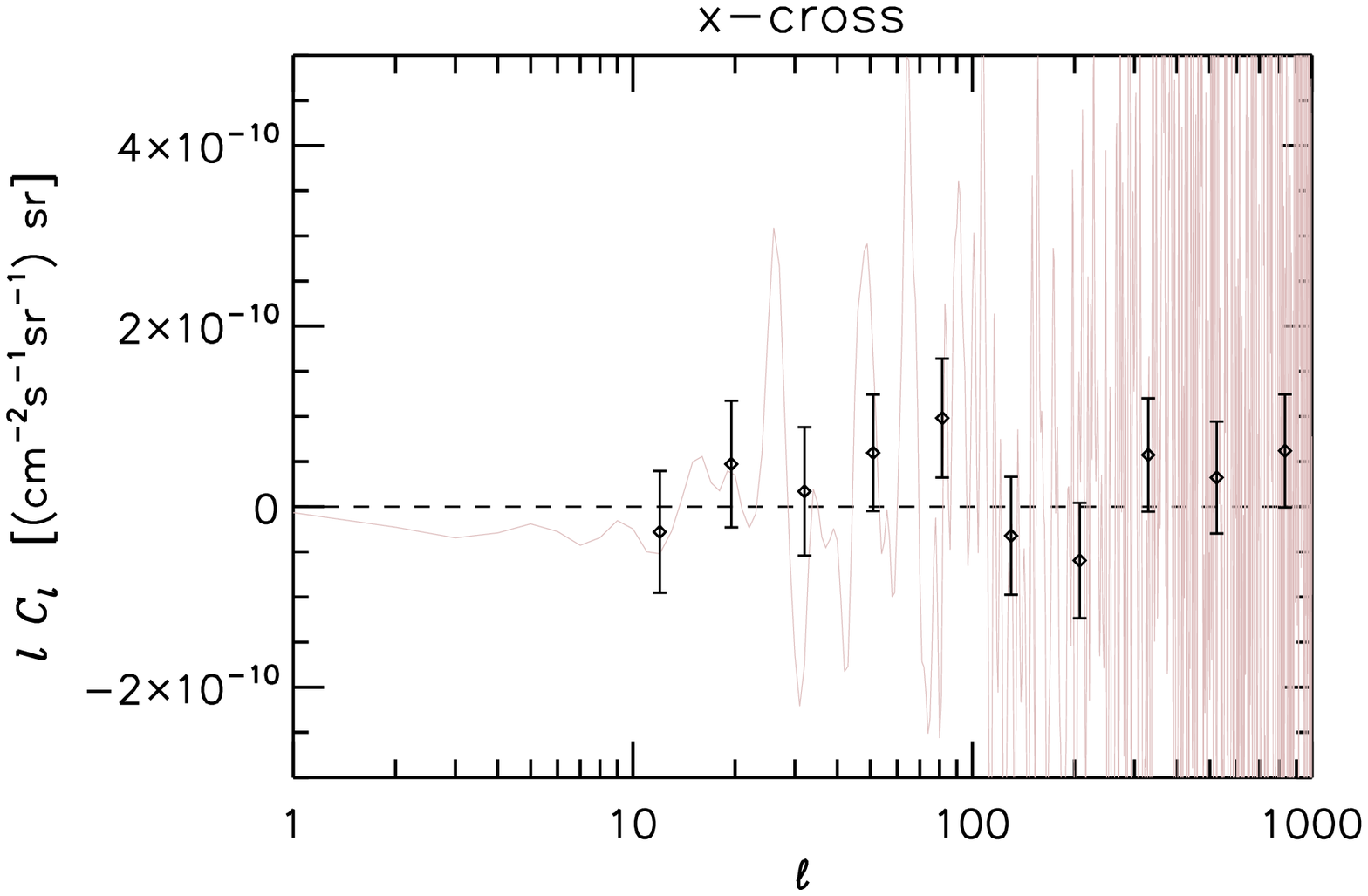}
\\ 
\includegraphics[width=.48\textwidth]{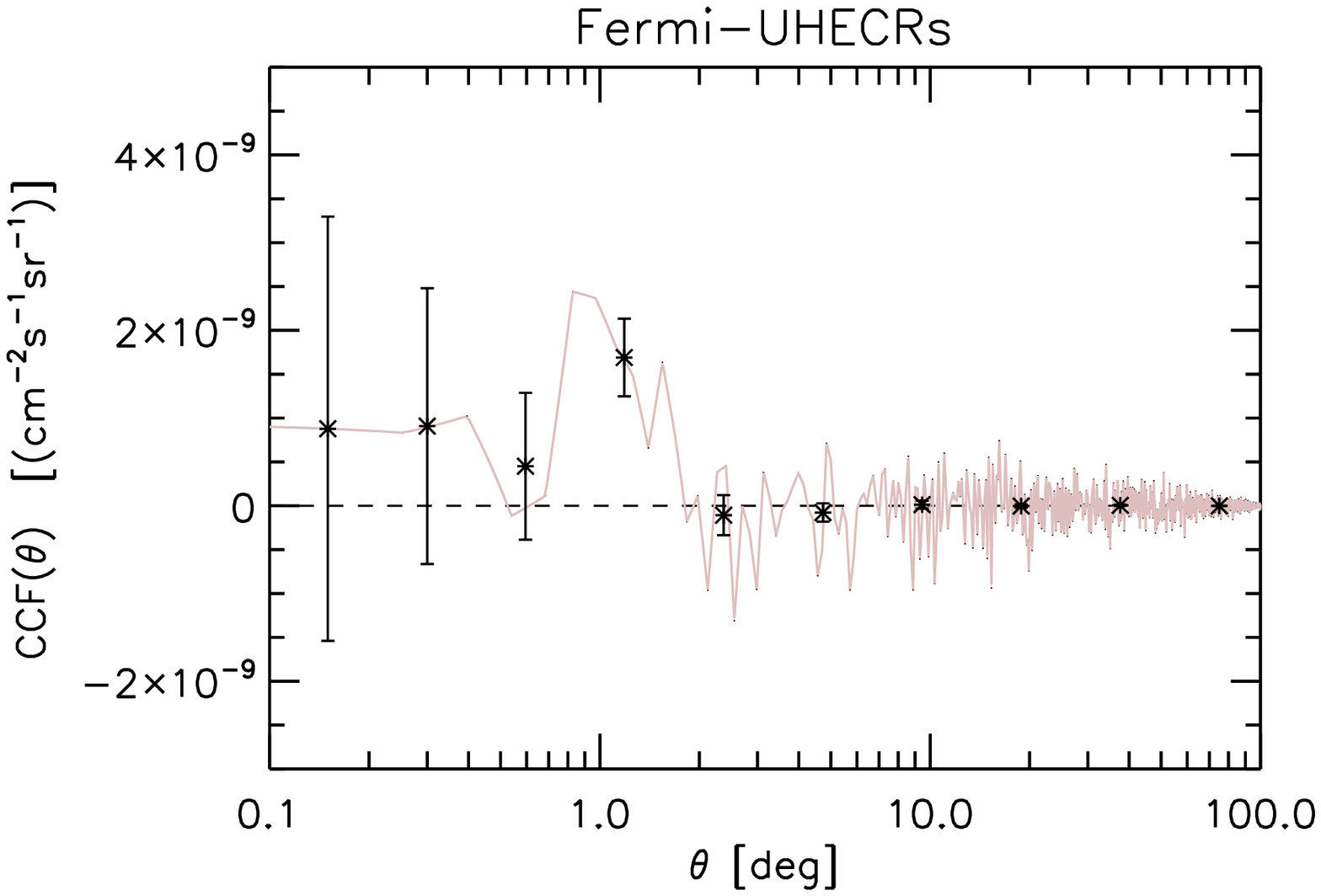}
\includegraphics[width=.48\textwidth]{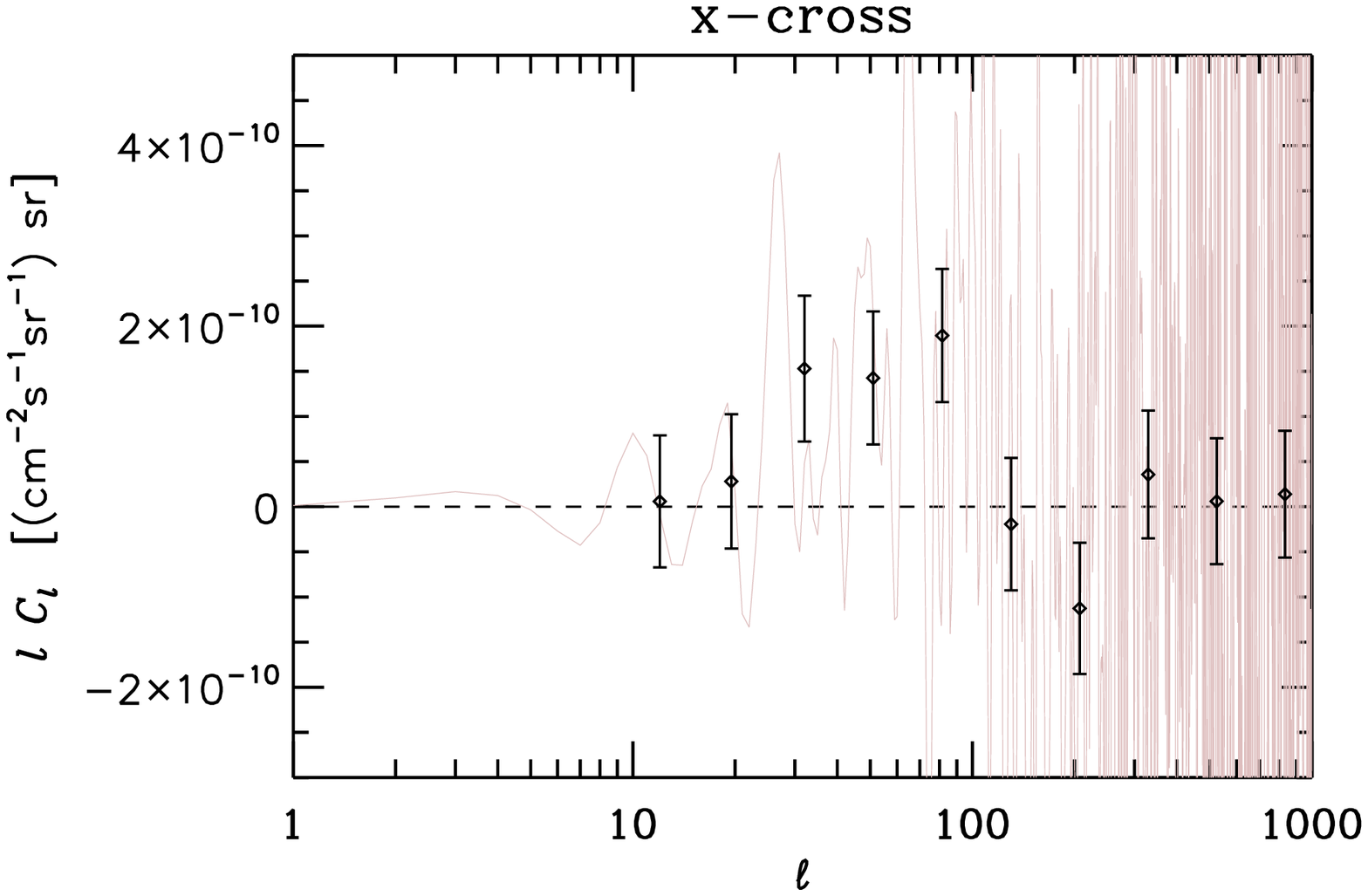}
\\ 
\includegraphics[width=.48\textwidth]{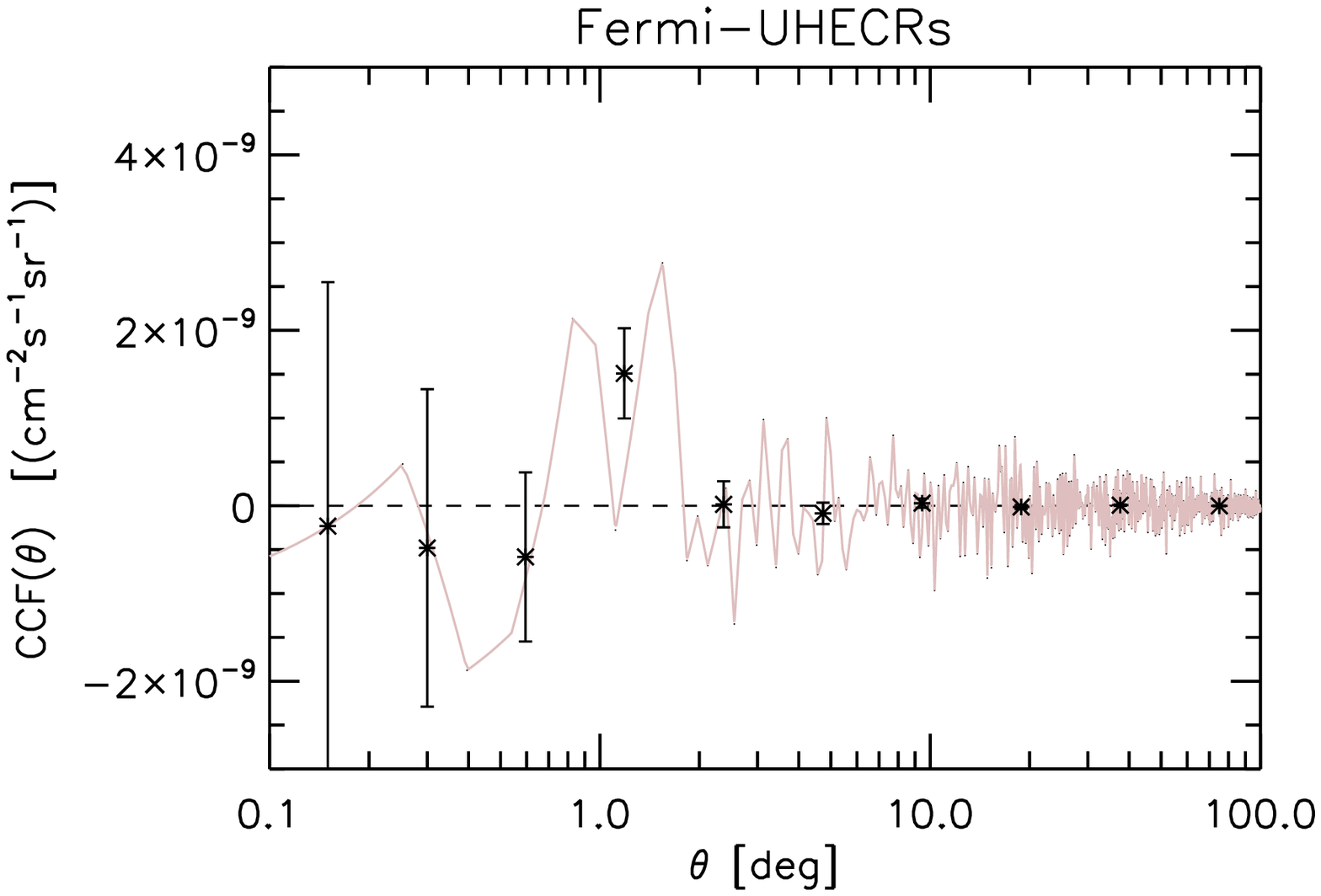}
\includegraphics[width=.48\textwidth]{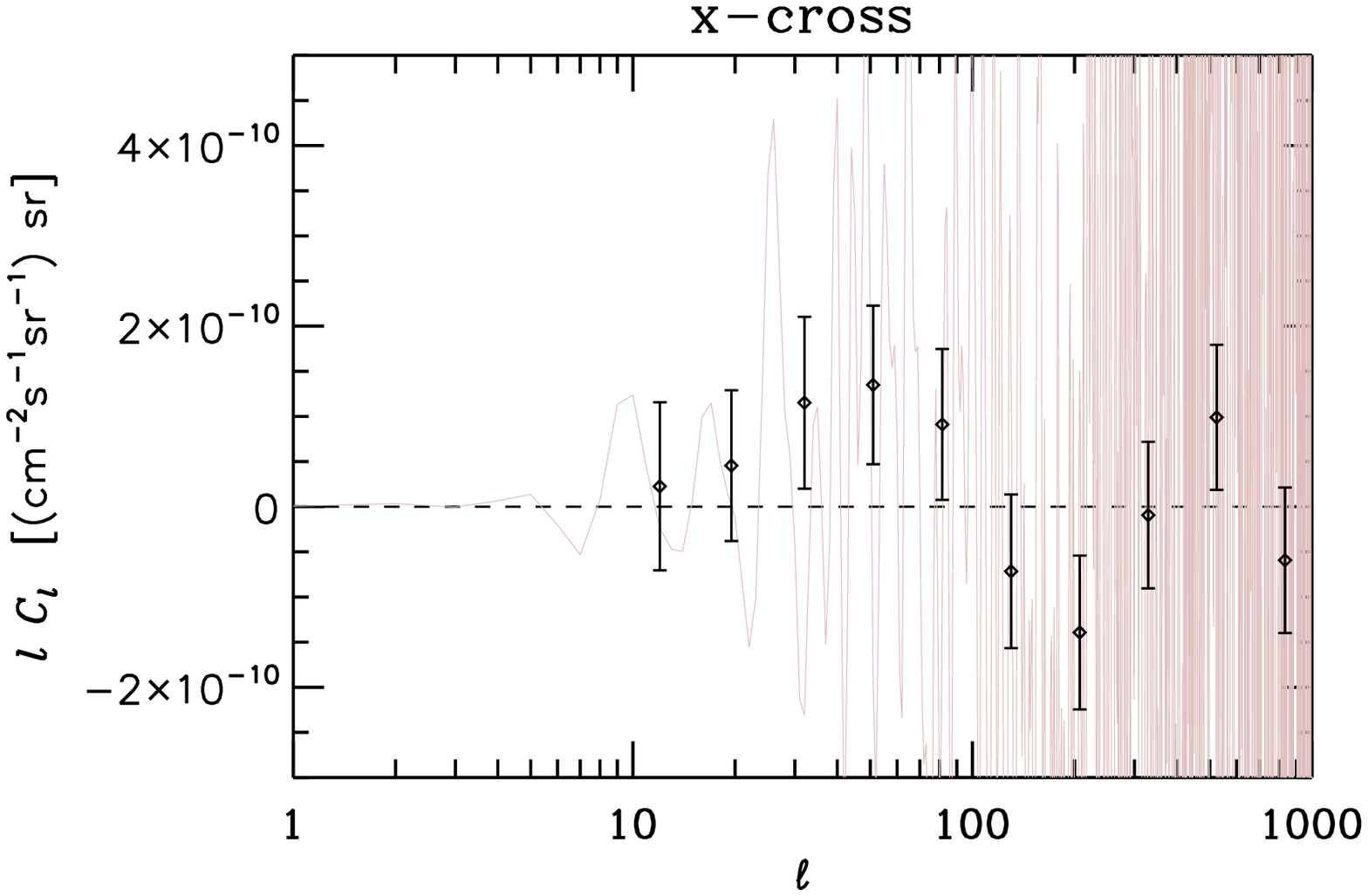}
\\ 
\vspace{-0.2cm}
\caption{Cross-correlation functions (left column) and correlation angular power spectra (right column) between 
the diffuse gamma-ray emission for $E>50$ GeV and UHECR directions.
Upper panels are for a Galactic latitude cut of $|b|>20^\circ$, middle panels for $|b|>30^\circ$ and lower panels for $|b|>40^\circ$.
In all panels the thin colored continuous line shows the unbinned angular spectra or correlation functions.}
\label{polspice50}
\end{figure}

Results are in Figures \ref{polspice1} and \ref{polspice50}.  For the case $E>1$ GeV we see that both CCF and CAPS are consistent with a null signal.
Above 50 GeV intriguingly there is an hint of correlation at an angular scale of about 1$^\circ$. The correlation is stronger
above 30$^\circ$ where the Galactic background is less intense.   Also, there is no correlation at scale smaller than 1$^\circ$
which would be consistent with the interpretation in terms of UHECRs deflected by about 1$^\circ$ from their original sources.
If this correlation is real, we expect it to  become more significant at even higher energies than 50 GeV,
becasue the gamma-ray horizon shrinks with increasing  energy matching better the UHECRs horizon.
We thus tested also for correlation above 200 GeV. To this purpose,  since the effective
area of P7REP data is not very large above 200 GeV giving poor statistics of photon events,
we use the newly available Pass 8\_Clean data selection, which is, instead, fully efficient in detecting photons
with energies up to $\sim$ 1 TeV.
Further, we use data spanning a period
of about 6 years, as opposed to the 5 year P7REP data-set. 
No significant correlation is, however, observed in this case, leaving inconclusive evidence for the hint above 50 GeV.
Also, the correlation above 50 GeV with Pass8 data remains similar, not showing an increase in significance with
respect to the P7REP case.
An improved dataset with more UHECRs and more gamma rays will help clarify in the future if the feature is real or a statistical fluctuation.

Given this slight excess in the Polspice analysis, we have re-done the calculations of 
Section \ref{ring} keeping $|b|>30^\circ$ and using as signal region $0-1.5^\circ$ and background region $4^\circ-5.5^\circ$, 
and we have not found any excess.  
We have also tested the case defining as signal region the angular range between $0.85^\circ$ and $1.5^\circ$, 
observing, again, no significant excess.

\section{Conclusions}\label{conclusions}
We have presented a search for correlation between high-energy (E$>1$ GeV) 
gamma-ray events from  {\it Fermi}-LAT and
UHECRs combining data sets from the Telescope Array
and the Pierre Auger Observatory above 75 EeV. We find no significant 
cross-correlation between UHECRs and 2FHL point sources.
There is also no hint of a diffuse gamma-ray excess  over the isotropic diffuse 
gamma-ray background in the cumulative data along the arrival directions
of our UHECR sample,
{\rev or in the direction of the TA and Cen A hot spots. In terms of the flux sensitivity above 10 GeV, we find the
2$\sigma$ flux to be $\sim 4 \cdot 10^{-11}$ ph cm$^{-2}$s$^{-1}$ for a
spectral index of $1.7$, comparable with the 1FHL  point-source sensitivity above 10  GeV
of   $\sim 5 \cdot  10^{-11}$ ph cm$^{-2}$s$^{-1}$.}
An increase of observed UHECRs and the continuation of the
{\it Fermi} mission should help collect additional data to 
improve existing limits.  With more events, it might be possible to
resolve the UHECR puzzle over the next decade.  

\section*{Acknowledgments}

The \textit{Fermi} LAT Collaboration acknowledges generous ongoing support
from a number of agencies and institutes that have supported both the
development and the operation of the LAT as well as scientific data analysis.
These include the National Aeronautics and Space Administration and the
Department of Energy in the United States, the Commissariat \`a l'Energie Atomique
and the Centre National de la Recherche Scientifique / Institut National de Physique
Nucl\'eaire et de Physique des Particules in France, the Agenzia Spaziale Italiana
and the Istituto Nazionale di Fisica Nucleare in Italy, the Ministry of Education,
Culture, Sports, Science and Technology (MEXT), High Energy Accelerator Research
Organization (KEK) and Japan Aerospace Exploration Agency (JAXA) in Japan, and
the K.~A.~Wallenberg Foundation, the Swedish Research Council and the
Swedish National Space Board in Sweden.
 
Additional support for science analysis during the operations phase is gratefully acknowledged from the Istituto Nazionale di Astrofisica in Italy and the Centre National d'\'Etudes Spatiales in France.

Authors are grateful to Giacomo Bonnoli 
for comments on the work. E.A.~acknowledges partial supported by ANPCyT PICT 2011-0359 and Asociaci\'on Universitaria Iberoam\'ericana de Postgrado. E.A. and G.Z. are grateful to ICTP for hospitality during development of part of the project.
This research was supported by a senior 
appointment to the NASA Postdoctoral Program
at the Goddard Space Flight Center, administered by 
Universities Space Research Association through a contract with NASA.


\begingroup\raggedright\endgroup

\end{document}